\documentclass[prd, aps, twocolumn, showpacs,superscriptaddress,nofootinbib]{revtex4-1}

\usepackage{amsbsy,amsmath,amssymb,amsthm,wasysym,amsfonts}
\usepackage{graphicx}
\usepackage{psfrag}
\usepackage{epstopdf}
\usepackage{color}
\usepackage{mathrsfs}
\usepackage{comment}
\usepackage[normalem]{ulem}
\DeclareSymbolFontAlphabet{\mathrsfs}{rsfs}
\DeclareMathAlphabet\mathbfcal{OMS}{cmsy}{b}{n}

\newcommand{\be}{\begin{equation}}  
\newcommand{\ee}{\end{equation}}
\newcommand{\bea}{\begin{eqnarray}}           
\newcommand{\eea}{\end{eqnarray}} 
\newcommand{\beqn}{\begin{eqnarray*}}
\newcommand{\eeqn}{\end{eqnarray*}}
\newcommand{\ba}{\begin{align}}
\newcommand{\ea}{\end{align}}

\def\lm{{\ell m}}   

\def\ii{{\rm i}}

\def\ha{{\hat{a}}}
\def\ta{{\tilde{a}}}

\def\bh{{\bar{h}}}

\def\p4{{\psi_4}} 

\usepackage{float}
\definecolor{cyan}{rgb}{0,0.9,0.9}
\definecolor{orange}{rgb}{0.9,0.5,0}
\definecolor{magenta}{rgb}{1,0,1}
\definecolor{purple}{rgb}{0.8,0.4,0.8}

\begin{document}


\title{An analytic family of post-merger template waveforms}

\author{Walter \surname{Del Pozzo}}
\affiliation{University of Birmingham, Edgbaston, B15 2TT
  Birmingham, United Kingdom}
\affiliation{Dipartimento di Fisica ``Enrico Fermi'', Universit\`a di Pisa, Pisa I-56127, Italy}
\author{Alessandro \surname{Nagar}}
\affiliation{Centro Fermi, Piazza del Viminale 1, 00184 Roma, Italy}
  \affiliation{INFN Sezione di Torino, Via P.~Giuria 1, 10125 Torino, Italy}
  \affiliation{Institut des Hautes Etudes Scientifiques, 91440
  Bures-sur-Yvette, France}

\begin{abstract}
Building on the analytical description of the post-merger (ringdown) waveform of coalescing,
non-precessing, spinning, (BBHs) introduced in Phys.~Rev.~D~90, 024054 (2014), 
we propose an analytic, closed form, time-domain, representation of the $\ell=m=2$ 
gravitational radiation mode emitted after merger. This expression is given as a function of the 
component masses and dimensionless spins $(m_{1,2},\chi_{1,2})$ of the two inspiralling objects, 
as well as  of the mass $M_{\rm BH}$ and (complex) frequency $\sigma_{1}$ of the fundamental 
quasi-normal mode of the remnant black hole. Our proposed template is obtained by fitting the 
post-merger waveform part of several publicly available numerical relativity simulations from the 
Simulating eXtreme Spacetimes (SXS) catalog and then suitably interpolating over (symmetric) 
mass ratio and spins. We show that this analytic expression accurately reproduces ($\sim$~0.01 rad) 
the phasing of the post-merger data of other datasets not used in its construction. This is notably 
the case of the spin-aligned run SXS:BBH:0305, whose intrinsic parameters are consistent 
with the 90\% credible intervals reported by the parameter-estimation followup of 
GW150914 in Phys.~Rev.~Lett.~116 (2016) no.24, 241102. Using SXS waveforms as ``experimental'' 
data, we further show that our template could be used on the actual GW150914 data to perform a 
new measure the complex frequency of the fundamental quasi-normal mode so to exploit the complete 
(high signal-to-noise-ratio) post-merger waveform. We assess the usefulness of our proposed template 
by analysing, in a realistic setting, SXS full inspiral-merger-ringdown waveforms and constructing
posterior probability distribution functions for the central frequency damping time of the first
overtone of the fundamental quasi-normal mode as well as for the physical parameters of the systems.
We also briefly explore the possibility opened by our waveform model to test the second law of
black hole dynamics. Our model will help improve current tests of general relativity,
in particular the general-relativistic no-hair theorem, and allow for novel tests,
such as that of the area theorem.
\end{abstract}

\date{\today}

\pacs{
   04.30.Db,  
    04.25.Nx,  
    95.30.Sf,  
 }

\maketitle

\section{Introduction}
One of the most difficult challenges provided by the discovery of  GW150914~\cite{Abbott:2016blz}
is to find robust evidence that the system is made by two black holes that eventually coalesce into a 
final black hole according to the predictions of general relativity (GR). An avenue pursued in ~\cite{TheLIGOScientific:2016src}
is to show consistency between the early-inspiral and the late-inspiral parts of the gravitational wave signal, 
with this latter, according to GR, dominated by the quasi-normal mode (QNM) frequencies of the final black hole. 
This nontrivial issue was carefully addressed in Ref.~\cite{TheLIGOScientific:2016src}, 
that followed the discovery paper~\cite{Abbott:2016blz}. In addition to the global waveform consistency test, 
Ref.~\cite{TheLIGOScientific:2016src} looked for evidence of the existence of the 
fundamental QNM in the post-merger signal. Ref.~\cite{TheLIGOScientific:2016src}
fit the signal at time $t>t_0$, with $t_{0}$ an arbitrary time after the merger time, $t_{\rm M}$, 
with a simple exponentially-damped oscillatory template of the form $Ae^{-\alpha (t-t_{0})}\cos[2\pi f_{0}(t-t_{0})+\phi_{0}]$. 
With this method, it was found that the 90\% posterior contour starts overlapping the 
GR prediction at $t_{0}=t_{\rm M}+3~{\rm ms}$, i.e. approximately $10M$ after the merger
point (see Fig.~4 in~\cite{TheLIGOScientific:2016src}). At later times, where theory 
predicts the fundamental QNM frequency to be persistent and dominant, the 
signal-to-noise-ratio (SNR) becomes too small, the statistical uncertainty becomes 
large and the signal is undetectable around $t_{0}\geq t_{\rm M}+8$~ms. 

The direct measurement of the frequency and damping time of the fundamental QNM of
the final black hole provides the most convincing evidence that the observed event is 
fully consistent with a binary black hole coalescence to a final single black hole as 
predicted by GR.  However, due to the nature of the fitting template mentioned above,
there are about $10M$ of signal, approximately corresponding to half a gravitational-wave 
(GW) cycle, which contributes significantly to the SNR, whose physical content is unexploited 
for the aforementioned analysis. Hence, the conclusions of Ref.~\cite{TheLIGOScientific:2016src} 
could be further strengthened by a reliable, analytical model, in the time-domain, 
of the complete {\it post-merger} part of the waveform (where the ``merger'' is
defined as the peak of the $|{\cal R }h_{22}|$ waveform) to be used as a fitting template.
This would allow for more flexibility in performing the analysis, that is currently 
limited, due to the particular choice of the fitting template, to just the late-time (low SNR) 
part of the signal, where among all QNMs (that get excited at the moment of merger), 
only the least-damped fundamental one is still present.
This paper introduces a novel waveform template, in the time-domain, 
designed to describe the {\it complete} post-merger signal, with the long term goal to
improve the post-merger analysis of GW150914 of Ref.~\cite{TheLIGOScientific:2016src} 
and similar signals that will likely be detected in the future by the LIGO and Virgo Collaborations. 
This template is based upon the analytical representation of the post-merger 
waveform for coalescing, non-precessing, BBH of Ref.~\cite{Damour:2014yha}.
This representation is obtained by interpolation of the primary fits of the post-merger 
numerical relativity (NR) waveform part after that the first, least-damped, QNM is 
factored out. The primary fit effectively models the presence of all the higher QNMs,
that are characterized by lower frequencies and shorter damping times than the 
fundamental one. Ref.~\cite{Damour:2014yha} focused on the equal-mass, equal-spin case 
only and thus used only the corresponding subset of the Simulating eXtreme Spacetimes (SXS)~\cite{sxs:catalog} 
catalog of numerical waveform data. All SXS waveforms were obtained with the 
Spectral Einstein Code~\cite{Chu:2009md,Mroue:2012kv,Hemberger:2013hsa,Lovelace:2010ne,Lovelace:2011nu,Buchman:2012dw,Mroue:2013xna,Scheel:2014ina}.
We generalize here the interpolating expressions  of Ref.~\cite{Damour:2014yha}, 
by including several of the unequal-mass, unequal-spin dataset present in the SXS catalog,
i.e. the waveform previously used for EOB/NR information and comparison in
Ref.~\cite{Nagar:2015xqa} plus a few more that were publicly available in June~2016,
when the first draft of this study was conceived, but we do not include
the dataset added to the catalog on October~31st, 2016 (see below).
We thus build a general analytical expression of the post-merger waveform that is 
a function of the symmetric mass ratio $\nu\equiv m_1 m_2/(m_1+m_2)^2$ and of 
the dimensionless spins $\chi_{1,2}\equiv S_{1,2}/(m_{1,2})^2$ of the two black holes as well as of
the final mass $M_{\rm BH}$ and (complex) frequency $\sigma_{1}$ of the fundamental QNM of the 
final remnant. Although we restrict, for simplicity, to considering only the $\ell=m=2$ mode, 
the method discussed here may be extended to model the post-merger part of 
subdominant multipolar modes~\footnote{This might be more complicated for modes like
  the $(3,2)$ that present mode-mixings due to the fact that the waveform is usually
  written as a multipolar decomposition over spin-weighted spherical harmonics.
  Future work may explore how the procedure discussed here could be applicable, for
  example on the waveform written using the basis of spheroidal harmonics~\cite{Kelly:2012nd}}.
The interpolating, improved, fit presented here is also now part of the NR-informed
${\tt EOB\_{ihes}}$ EOB code~\cite{Nagar:2015xqa,Damour:2014sva}. 

The paper is organized as follows: In Sec.~\ref{sec:build_template} we construct the analytic
template waveform, while Sec.~\ref{sec:accuracy} is devoted to testing is accuracy and
reliability. The performance of the template in a simulated data-analysis setup is
evaluated in Sec.~\ref{sec:da_stuff} and we summarize our findings in Sec.~\ref{sec:conclusions}.

\section{Template construction}
\label{sec:build_template}
We begin by introducing a convenient notation. The multipolar decomposition of the 
waveform is given by $ h_+-{\rm i} h_\times = \sum_{\ell,m}h_{\ell m} {}_{-2}Y_\lm(\theta,\phi)$,
and we focus on the $\ell=m=2$ ``post-merger'', $\nu$-scaled, 
waveform,
\be\label{eq:wf}
h(\tau)\equiv \dfrac{1}{\nu}\dfrac{{\cal R} c^2}{GM}h_{22}^{\rm postmerger}(\tau),
\ee
where $M\equiv m_1+m_2$ is the total mass and ${\cal R}$ is the distance of the source.
The time $\tau=(t-t_{\rm M})/M_{\rm BH}$ counts time in units of the mass
of the final black hole, $M_{\rm BH}$, and $t_{\rm M}$ is the merger time.
The {\it QNM-rescaled ringdown waveform} $\bar{h}(\tau)$ of~\cite{Damour:2014yha} 
$h(\tau)$ is defined as $h(\tau)\equiv e^{-\sigma_1 \tau - {\rm i}\phi_0}\bar{h}(\tau)$,
where $\sigma_1\equiv \alpha_1+{\rm i}\omega_1$ is the (dimensionless, $M_{\rm BH}$-rescaled)
complex frequency of the fundamental (positive frequency, $\omega_1>0$) QNM of
the final black hole and $\phi_0$ is the value of the phase at merger.
The (complex) function $\bar{h}(\tau)$ is then decomposed in amplitude and phase
as
\be
\label{eq:barh}
\bar{h}(\tau)\equiv A_{\bar h}e^{{\rm i}\phi_{\bh}(\tau)}.
\ee
Reference~\cite{Damour:2014yha} found that  $A_{\bh}$ and $\phi_{\bh}$
can be accurately represented by the following general functional forms
\begin{align}
\label{eq:barA}
A_\bh(\tau)    &=c_1^A {\rm tanh}(c_2^A\tau + c_3^A)+c^A_4, \\
\label{eq:barPhi}
\phi_\bh(\tau) &=-c_1^\phi\ln\left(\dfrac{1+c_3^\phi e^{-c_2^\phi \tau}+c_4^\phi e^{-2c_2^\phi\tau}}{1+c_3^\phi+ c_4^\phi}\right).
\end{align} 
\begin{table}[t]
\caption{The $\nu$-dependence of the coefficients in Eq.~\eqref{eq:Yvsa0}.}
\begin{center}
\begin{ruledtabular}
\begin{tabular}{cccc}

$A_{\alpha_{21}}= $&$-0.0185533\,\nu$ &$-0.0166417 $&\\
$B_{\alpha_{21}}= $&$-0.0594092\,\nu$ &$-0.0157896 $&\\
$C_{\alpha_{21}}= $&$-0.100191\,\nu$ &$+0.19044 $&\\
\hline

$A_{\alpha_{1}}= $&$-0.0123998\,\nu$ &$-0.00791069 $&\\
$B_{\alpha_{1}}= $&$-0.0421559\,\nu$ &$-0.00365094 $&\\
$C_{\alpha_{1}}= $&$-0.040671\,\nu$ &$+0.0919055 $&\\
\hline

$A_{c_{3}^{A}}= $&$+0.417778\,\nu$ &$-0.0175206 $&\\
$B_{c_{3}^{A}}= $&$+0.0243799\,\nu$ &$-0.22621 $&\\
$C_{c_{3}^{A}}= $&$+0.953089\,\nu$ &$-0.592121 $&\\
\hline
$A_{c_{3}^{\phi}}= $&$+12.9727\,\nu$ &$-0.350191 $&\\
$B_{c_{3}^{\phi}}= $&$-0.249142\,\nu$ &$+3.10091 $&\\
$C_{c_{3}^{\phi}}= $&$-1.6901\,\nu$ &$+4.44107 $&\\
\hline
$A_{c_{4}^{\phi}}= $&$+23.3553\,\nu$ &$+1.9222 $&\\
$B_{c_{4}^{\phi}}= $&$-0.448352\,\nu$ &$+4.99249 $&\\
$C_{c_{4}^{\phi}}= $&$-3.05867\,\nu$ &$+2.70508 $&\\
\hline
$A_{\Delta\omega}= $&$+0.129442\,\nu$ &$+0.0232987 $&\\
$B_{\Delta\omega}= $&$+0.165507\,\nu$ &$+0.0517482 $&\\
$C_{\Delta\omega}= $&$+0.383848\,\nu$ &$+0.0850474 $&\\
\hline
$A_{\hat{A}_{22}^{\rm mrg}}= $&$+0.229867\,\nu$ &$-0.0411679 $&\\
$B_{\hat{A}_{22}^{\rm mrg}}= $&$-0.450254\,\nu$ &$+0.107428 $&\\
$C_{\hat{A}_{22}^{\rm mrg}}= $&$+0.742481\,\nu$ &$+1.38748 $&\\
\hline

$A_{\omega_{22}^{\rm mrg}}= $&$-0.285624\,\nu$ &$+0.0903558 $&\\
$B_{\omega_{22}^{\rm mrg}}= $&$-0.185274\,\nu$ &$+0.12597 $&\\
$C_{\omega_{22}^{\rm mrg}}= $&$+0.405274\,\nu$ &$+0.258643 $& \\
\end{tabular}
\end{ruledtabular}
\end{center}
\label{tab:fitting}
\end{table}
As in Ref.~\cite{Damour:2014yha}, only three of the eight fitting coefficients, 
$(c_{3}^{A},c_{3}^{\phi},c_{4}^{\phi})$, are independent 
and need to be fit directly. The others can be expressed in terms of 
four other physical quantities:  ($\alpha_{1},\alpha_{21},\Delta\omega,\hat{A}_{22}^{\rm mrg}$)
 \begin{align}
\label{eq:c2A}
 c_{2}^{A}     & = \dfrac{1}{2}\alpha_{21},\\
\label{eq:c4A}
 c_{4}^{A}     & = \hat{A}_{22}^{\rm mrg}-c_{1}^{A}\tanh( c_{3}^{A}),\\
\label{eq:c1A}
 c_{1}^{A}     & = \hat{A}_{22}^{\rm mrg}\,\alpha_{1}\dfrac{\cosh^{2}(c_{3}^{A})}{c_{2}^{A}},\\
\label{eq:c1phi}
 c_{1}^{\phi} & = \Delta\omega\dfrac{1+c_{3}^{\phi}+c_{4}^{\phi}}{c_{2}^{\phi}(c_{3}^{\phi}+2c_{4}^{\phi})},\\
\label{eq:c2phi}
 c_{2}^{\phi} & = \alpha_{21},
 \end{align}
 
\begin{table*}[t]
\caption{Dataset of the SXS catalog used for the cross-validation of the template waveform, see Fig.~\ref{fig:figFit}. 
  The datasets marked with an * were used in the construction of the template. The dataset SXS:BBH:none is a 14~orbit
  waveform not part of the SXS catalog~\cite{Buchman:2012dw} that was used to calibrate EOB models~\cite{Pan:2011gk,Damour:2012ky}. 
  From left to right, the columns report: the dataset number in the SXS catalog, the mass ratio $q=m_1/m_2$;
  the symmetric-mass ratio $\nu\equiv m_1 m_2/(m_1+m_2)^2$;
  the dimensionless spins; the mass and angular momentum of the final black hole $(M_{\rm BH},J_{\rm BH})$; the NR phase
  at merger time ($\tau=0$), $\phi_0^{\rm NR_{mrg}}$; and the corresponding GW frequency $M_{\rm BH}\omega_{22}^{\rm mrg}$. The value of $\phi_0^{\rm NR_{\rm mrg}}$ is used to align the NR and analytical
    waveforms at $\tau=0$ in Fig.~\ref{fig:alignatmrg}.}
\begin{center}
\begin{ruledtabular}
\begin{tabular}{ccccccccc}
ID  & $q$ &$\nu$ & $S_{1}/(m_{1})^{2}$ & $S_{2}/(m_{2})^{2}$ & $M_{\rm BH}/M$& $J_{\rm BH}/M_{\rm BH}^{2}$ &$\phi_0^{\rm NR_{mrg}}$ & $M_{\rm BH}\omega_{22}^{\rm mrg}$\\
\hline
SXS:BBH:none*     &  1     &  $0.25$         & $0$        & $0$         & 0.95161 & 0.6864      &  $0.8579$  & 0.3422   \\ 
SXS:BBH:0152*  &  1     &  $0.25$         &  $+0.60$   & $+0.60$     & 0.9269  & 0.8578      & $-2.7441$ & 0.3849   \\ 
SXS:BBH:0211   &  1     &  $0.25$         &  $+0.90$   & $-0.90$     &  0.9511 & 0.6835      & $2.1415$  & 0.3415   \\
SXS:BBH:0178*  &  1     &  $0.25$         & $+0.994$   & $+0.994$    &  0.8867 & 0.9499      & $0.0856$  & 0.4199   \\
SXS:BBH:0305   & 1.221  & $0.2475$        & $+0.33$    & $-0.4399$   & 0.9520  & 0.6921      & $ 2.6594$ & 0.3427   \\ 
SXS:BBH:0025   & 1.5    & $0.2400$        & $+0.4995$  & $-0.4995$   & 0.9504  & 0.7384      & $-0.4623$ & $0.3513$ \\ 
SXS:BBH:0184   & 2      & $0.\bar{2}$     & $0$        & 0           & 0.9612  & 0.6234      & $-3.0966$ & 0.3336   \\ 
SXS:BBH:0162   & 2      & $0.\bar{2}$     & $+0.60$    & $0$         & 0.9461  & 0.8082      & $-1.6577$ & 0.3687   \\ 
SXS:BBH:0257   & 2      & $0.\bar{2}$     & $+0.85$    & $+0.85$     & 0.9199  & 0.9175      & $2.2626$  & 0.4152   \\
SXS:BBH:0045   & 3      & $0.1875$        & $+0.4995$  & $-0.4995$   & 0.9628  & 0.7410      & $-2.6842$ & 0.3617   \\ 
SXS:BBH:0292   & 3      & $0.1875$        &  $+0.7314$ & $-0.8493$   & 0.9560  & 0.8266      & $1.9684$  & 0.3750   \\
SXS:BBH:0293   & 3      & $0.1875$        & $+0.85$    & $+0.85$     & 0.9142  & 0.9362      & $-2.6663$ & $0.4158$ \\ 
SXS:BBH:0317   &  3.327 &  $0.1777$       &  $0.5226$  & $-0.4482$   & 0.9642  & 0.7462      & $-0.3756$ & 0.3677   \\ 
SXS:BBH:0208*  &    5   &   $0.13\bar{8}$ &  $-0.90$   & 0           & 0.98822 & $-0.12817$  & $+0.5148$ & 0.2626   \\
SXS:BBH:0203   &  7     &  $0.1094$       & $+0.40$    & $0$         & 0.9836  & $0.6056$    & $-0.9013$ & 0.3341   \\                  
SXS:BBH:0207   &  7     &  $0.1094$       & $-0.60$    & $0$         & 0.9909  & $-0.0769$   & $-1.3736$ & 0.2631   \\
SXS:BBH:0064*  & 8      & $0.0987$        & $-0.50$    & $0$         & 0.9922  & $-0.0526$   & $2.3926$  & 0.2634   \\
SXS:BBH:0185 & 9.990    & $0.0827$        &  $0$       & $0$         & 0.9917  & $0.2608$    & $0.4982$  & 0.2948   
\end{tabular}
\end{ruledtabular}
\end{center}
\label{tab:NRtest}
\end{table*}

\begin{table*}[t]
  \caption{The last  two columns list fundamental QNMs frequencies and final black hole mass inferred from NR data and
    measured with the post-merger template, after adding to the NR waveform white Gaussian noise. For all NR waveforms,
    the total mass is fixed to $M=60M_\odot$ and we consider postmerger SNR=10 and SNR=50. The uncertainty on the measured
    quantities corresponds to the $90\%$ credible regions. The measures of both  $M_{\rm BH}$ and $\sigma_1$ are biased
    for some specific datasets with large mass ratio and spin anti-aligned with the orbital angular momentum.}
\begin{center}
\begin{ruledtabular}
\begin{tabular}{cc|c||cc|cl}
ID  & $(q,\chi_1,\chi_2)$ & $\sigma_{1}^{\rm NR}$&  $\sigma_{1}|_{\rm SNR=10}$ & $M_{\rm BH}/M_\odot|_{\rm SNR=10}$ & $\sigma_{1}|_{\rm SNR=50}$  & $M_{\rm BH}/M_{\odot}|_{\rm SNR=50}$\\
\hline
         SXS:BBH:none*     & $(1,0,0)$           & $0.0813 + \ii 0.527$           &  $0.07_{-0.02}^{+0.03}+\ii 0.52_{-0.17}^{+0.19}$ & $57.1_{-17.2}^{+19.6}$ & $0.08_{-0.01}^{+0.01}+\ii 0.50_{-0.06}^{+0.06}$ & $54.3_{-5.4}^{+5.4}$ \\
SXS:BBH:0152*  & $(1,+0.60,+0.60)$   & $0.0706+ \ii 0.629$ &  $0.07_{-0.03}^{+0.03}+\ii 0.78_{-0.31}^{+0.20}$ & $69.5_{-27.2}^{+18.0}$ & $0.08_{-0.01}^{+0.01}+\ii 0.67_{-0.11}^{+0.10}$ & $59.0_{-8.6}^{+7.6}$\\     
SXS:BBH:0211   & $(1,-0.90,+0.90)$   & $0.081+  \ii 0.525$ & $0.11_{-0.04}^{+0.04}+\ii 0.85_{-0.33}^{+0.14}$ & $99.6_{-38.0}^{+17.7}$  & $0.08_{-0.01}^{+0.02}+\ii 0.50_{-0.08}^{+0.09}$ & $56.2_{-7.5}^{+8.6}$\\               
SXS:BBH:0178*  & $(1,+0.994,+0.994)$ & $0.053 + \ii0.746$  &$0.08_{-0.03}^{+0.04}+\ii 0.74_{-0.23}^{+0.22}$ & $55.3_{-16.2}^{+17.1}$  & $0.06_{-0.00}^{+0.01}+\ii 0.73_{-0.04}^{+0.05}$ & $52.7_{-2.4}^{+3.1}$\\                    
SXS:BBH:0305   & $(1.2,+0.33,-0.44)$ & $0.081 + \ii0.529$ & $0.11_{-0.05}^{+0.09}+\ii 0.88_{-0.33}^{+0.11}$ & $92.4_{-34.3}^{+15.5}$  & $0.10_{-0.03}^{+0.04}+\ii 0.62_{-0.18}^{+0.26}$ & $65.7_{-17.6}^{+24.8}$\\
SXS:BBH:0025   & $(1.5,+0.5,-0.5)$   & $0.079 + \ii 0.550$ & $0.12_{-0.04}^{+0.06}+\ii 0.90_{-0.27}^{+0.09}$ & $94.8_{-28.3}^{+13.1}$ & $0.09_{-0.02}^{+0.02}+\ii 0.58_{-0.12}^{+0.14}$ & $60.6_{-11.0}^{+12.7}$\\
SXS:BBH:0184   & $(2,0,0)$           & $0.083 + \ii 0.502 $& $0.26_{-0.22}^{+0.22}+\ii 0.54_{-0.41}^{+0.41}$ & $39.0_{-28.2}^{+116.4}$& $0.13_{-0.06}^{+0.09}+\ii 0.85_{-0.34}^{+0.14}$ & $97.8_{-40.1}^{+22.4}$\\
SXS:BBH:0162   & $(2,+0.60,0)$       & $0.075 + \ii 0.591 $ & $0.11_{-0.06}^{+0.08}+\ii 0.74_{-0.35}^{+0.24}$ & $77.5_{-37.1}^{+29.0}$& $0.07_{-0.01}^{+0.01}+\ii 0.59_{-0.09}^{+0.10}$ & $56.4_{-7.5}^{+8.0}$\\
SXS:BBH:0257   & $(2,+0.85,+0.85)$   & $0.062 + \ii0.694$  & $0.08_{-0.04}^{+0.05}+\ii 0.74_{-0.29}^{+0.23}$ & $64.3_{-24.6}^{+21.6}$ & $0.07_{-0.01}^{+0.01}+\ii 0.71_{-0.06}^{+0.06}$ & $56.2_{-3.8}^{+4.1}$\\
SXS:BBH:0045   & $(3,+0.5,-0.5)$     & $0.079 + \ii 0.552$ & $0.31_{-0.23}^{+0.17}+\ii 0.53_{-0.40}^{+0.43}$ & $131.1_{-86.2}^{+52.8}$& $0.14_{-0.05}^{+0.05}+\ii 0.88_{-0.29}^{+0.11}$ & $89.9_{-28.7}^{+14.0}$\\
SXS:BBH:0292   & $(3,+0.73,-0.85)$   & $0.073 + \ii0.604$  &   $0.11_{-0.05}^{+0.05}+\ii 0.85_{-0.28}^{+0.14}$ & $86.2_{-29.7}^{+17.9}$ & $0.08_{-0.01}^{+0.01}+\ii 0.64_{-0.07}^{+0.08}$ & $60.1_{-5.7}^{+6.1}$\\
SXS:BBH:0293   & $(3,+0.85,+0.85)$   & $0.062 + \ii0.689$  &  $0.09_{-0.03}^{+0.04}+\ii 0.86_{-0.26}^{+0.13}$ & $74.2_{-22.0}^{+13.0}$ & $0.08_{-0.01}^{+0.01}+\ii 0.77_{-0.09}^{+0.11}$ & $62.6_{-6.0}^{+7.1}$ \\    
SXS:BBH:0317   & $(3.3,0.52,-0.45)$  & $0.078 + \ii 0.554$ & $0.08_{-0.04}^{+0.05}+\ii 0.71_{-0.33}^{+0.26}$ & $74.3_{-33.6}^{+26.8}$ & $0.08_{-0.01}^{+0.02}+\ii 0.56_{-0.09}^{+0.12}$ & $59.0_{-8.4}^{+11.8}$\\
SXS:BBH:0208*  & $(5,-0.90,0)$       & $0.089 + \ii0.359$  & $0.15_{-0.07}^{+0.07}+\ii 0.58_{-0.27}^{+0.33}$ & $84.9_{-38.3}^{+47.6}$& $0.05_{-0.00}^{+0.11}+\ii 0.21_{-0.02}^{+0.36}$ & $33.8_{-2.5}^{+55.6}$\\
SXS:BBH:0203   & $(7,+0.40,0)$       & $0.083 + \ii0.495$  & $0.07_{-0.03}^{+0.06}+\ii 0.48_{-0.18}^{+0.36}$ & $58.3_{-21.0}^{+41.4}$& $0.07_{-0.01}^{+0.01}+\ii 0.45_{-0.04}^{+0.04}$ & $54.1_{-4.3}^{+4.6}$\\
SXS:BBH:0207   & $(7,-0.60,0)$       & $0.089 + \ii0.364$  &  $0.10_{-0.05}^{+0.08}+\ii 0.43_{-0.21}^{+0.37}$ & $70.3_{-32.7}^{+58.8}$ & $0.08_{-0.01}^{+0.01}+\ii 0.33_{-0.03}^{+0.04}$ & $54.2_{-5.0}^{+6.0}$\\
SXS:BBH:0064*  & $(8,-0.50,0)$       & $0.089 + \ii 0.367$ & $0.15_{-0.09}^{+0.17}+\ii 0.84_{-0.41}^{+0.15}$ & $108.2_{-54.7}^{+36.4}$& $0.08_{-0.02}^{+0.03}+\ii 0.51_{-0.12}^{+0.18}$ & $58.7_{-13.3}^{+18.9}$\\
SXS:BBH:0185   & $(9.99,0,0)$        & $0.087 + \ii 0.412$ & $0.19_{-0.09}^{+0.12}+\ii 0.77_{-0.35}^{+0.21}$ & $108.9_{-51.8}^{+42.0}$ &  $0.13_{-0.03}^{+0.03}+\ii 0.59_{-0.12}^{+0.15}$ & $79.8_{-15.3}^{+18.7}$\\
\end{tabular}
\end{ruledtabular}
\end{center}
\label{tab:measures}
\end{table*}

because of physical constraints imposed on the template~\eqref{eq:barA}-\eqref{eq:barPhi} (see also Ref~\cite{Damour:2014yha}). 
Here $\alpha_{21}\equiv \alpha_{2}-\alpha_{1}$, where $\alpha_{2}$ is the inverse damping time of the 
first overtone of the fundamental quasi-normal-mode of the final black hole; $\hat{A}_{22}^{\rm mrg}\equiv |h(0)|$ 
is the $\nu$-rescaled waveform amplitude at merger, and 
finally $\Delta\omega\equiv \omega_{1}-M_{\rm BH}\omega_{22}^{\rm mrg}$, where
$\omega_{22}^{\rm mrg}$ is the GW frequency at merger. 
The quantities $(\alpha_{1},\alpha_{21},\Delta\omega,\hat{A}_{\rm mrg},\omega_{22}^{\rm mrg})$ are extracted directly 
from each SXS waveform data set (extrapolated with $N=3$ at infinite extraction radius~\cite{sxs:catalog}), 
notably using the information available in the {\tt metadata.txt} file coming with each data set 
(e.g., $M$, $M_{\rm BH}$ and $J_{\rm BH}$) to obtain the corresponding QNMs frequencies by 
interpolating the publicly available data from E.~Berti website~\cite{berti:qnms}.  
The other parameters, $(c_{3}^{A},c_{3}^{\phi},c_{4}^{\phi})$, are obtained by fitting the post-merger part 
($\tau\geq 0$) of $\bar{h}(\tau)$ with the fitting templates~\eqref{eq:barA}-\eqref{eq:barPhi} constrained 
by Eqs.~\eqref{eq:c2A}-\eqref{eq:c2phi}. The time interval over which the fit is done typically corresponds 
to four times the damping time of the first QNMs, i.e. $4/\alpha_{1}$ and it is limited by the spurious 
oscillations that occur in the numerical $\bar{h}(\tau)$ functions at later times~\cite{Damour:2014yha}.
Eventually, each SXS post-merger, QNM-scaled, waveform can be characterized by the vector 
$Y\equiv \left( \alpha_{1},\alpha_{21},c_{3}^{A},c_{3}^{\phi},c_{4}^{\phi},\hat{A}_{22}^{\rm mrg},\Delta\omega\right)$,
whose elements depend on the mass ratio and spins of the binary. 
To determine the general functional dependence of the vector $Y$ on the binary parameters, 
the equal-mass, equal-spin datasets used in Refs.~\cite{Damour:2014yha,Nagar:2015xqa}
are complemented by further SXS datasets with $(8,\pm 0.5,0)$, $(8,0,0)$, $(5,\pm 0.5,0)$,$(5,-0.9,0)$, $(5,0,0)$ 
$(3,\pm 0.5,0)$, $(3,\pm 0.5,\pm 0.5)$, $(3,0,0)$, $(1.5,\pm0.5,0)$, $(1.5,0,0)$.
The parameter space coverage of NR waveforms is rather scarce:
away from the equal-mass, equal-spin case (where one relies on over 20 NR simulations, 
see Table~I of Ref.~\cite{Nagar:2015xqa}), only few data points are available for the fitting procedure. 
The SXS catalog provides 3 data points for $\nu=6/25$ ($q=3/2$), 
5 points for $\nu=3/16$ ($q=3$), 4 points for $\nu=5/36$ ($q=5$) and 3 points for $\nu=8/81$ ($q=8$).
As a consequence, at each fixed value of $q$, the spin-dependence can be modeled with no 
more than three parameters~\footnote{While this paper was under review, on October 31st 2016, 
the SXS collaboration added to the catalog the 95 new simulations presented in~\cite{Chu:2015kft}.
This additional NR information was not used in the construction of the post-merger template, a work
that will be addressed in future studies; rather, we used a few of the datasets of Ref.~\cite{Chu:2015kft}
just to evaluate the accuracy of the model, as we shall discuss below.}.
The analytical representation of the post-merger waveform  as a function of $\nu$ and (some) spin variables is obtained with the following 3-step procedure: 
(i) for each configuration $(q,\chi_1,\chi_2)$ we obtain the vector $Y_{\nu,\chi_1,\chi_2}$; 
(ii) then, for each value of $\nu$, we fit $Y$ versus the dimensionless spin parameter 
$\hat{a}_{0}=X_{1}\chi_{1}+X_{2}\chi_{2}$ (where $X_{1,2}=m_{1,2}/M$), 
with a quadratic function of $\ha_{0}$
\begin{align}
\label{eq:Yvsa0}
Y(\nu,a_0) = A_{Y}(\nu)\ha_{0}^{2} + B_{Y}(\nu) \ha_{0} + C_Y(\nu)\,.
\end{align}
In general, we would expect $Y$ to be function of three variables 
$Y\equiv Y(\nu,\ta_1,\ta_2)$,  with $\ta_{1,2}\equiv X_{1,2}\chi_{1,2}$ and 
{\it not} of just the sum $\ha_0=\ta_1 + \ta_2$, and possibly not just
a quadratic function of the spin, \emph{e.g.}~\cite{Damour:2014yha}. 
A more complicated spin-dependence may be needed when more
(spin-aligned) NR simulations will be incorporated.
Finally, (iii) we found that linear functions in $\nu$ are
sufficient to model the vectors of coefficients,
$(A_{Y}(\nu),B_{Y}(\nu),C_{Y}(\nu))$. Eventually, each function composing the 
vector $Y(\nu,\ha_0)= (\alpha_{1},\alpha_{21},c_{3}^{A},c_{3}^{\phi},c_{4}^{\phi},\hat{A}_{22}^{\rm mrg},\Delta\omega)$
that defines the post-merger template is given by six numbers from Table~\ref{tab:fitting}.

Before evaluating quantitatively the performance of the global fit, let us briefly outline 
some of its limitations, that are present already at the level of the  primary fitting template 
of Eqs.~\eqref{eq:barA}-\eqref{eq:barPhi}. The most relevant drawback of our analytical 
ansatz is that it is unable, by construction, to take into account 
the behavior entailed by the simultaneous presence of retrograde ($m>0$)
and prograde  ($m<0$) QNMs. As it is well known 
(see e.g.~\cite{Damour:2007xr,Harms:2014dqa,Taracchini:2014zpa}
for the large-mass-ratio limit case, where the effect is maximal),
when $m<0$ modes are excited the waveform modulus and frequency show oscillations due to 
mode mixing, the amplitude of these oscillations mirroring the relative importance 
of the two QNMs branches. Such interference is more marked as the mass ratio 
increases and/or the spins are anti-aligned with the total angular momentum and large. 
Due to the absence of any oscillatory term in the
ansatzs~\eqref{eq:barA}-\eqref{eq:barPhi} for $(A_{\bar{h}},\phi_{\bar{h}})$, our
representation of $\bar{h}$ is, a priori, not expected to be faithful
in this case, which may eventually result in biases in the recovered
parameters. We shall briefly come back on this point below, although
a thorough analysis of these effects is postponed to future studies.

Similarly, one also finds that the primary fit might be inaccurate when applied
to the post-merger part of the large-mass-ratio waveforms of Ref.~\cite{Harms:2014dqa}
obtained with black hole spin large and {\it anti-aligned} with the orbital angular momentum,
due to the lack of modelization of mode-mixings effects. When the spin is similarly large,
but {\it aligned} with the orbital angular momentum, one also finds that
Eqs.~\eqref{eq:barA}-\eqref{eq:barPhi} are not sufficiently flexible to allow for
an accurate fit of the post-merger part, especially of the amplitude. One indeed
finds that the fit amplitude typically develops a secondary peak in the postmerger
region, and the amplitude of this secondary peak becomes larger as the spin gets close to 1.
In a preliminary study, we could see  that this behavior is qualitatively present
for mass ratios of order 3 or 5 and relatively mild spins (e.g. $+0.5$), although
the effect is small enough to be considered irrelevant. By contrast, when
the mass ratio increases, e.g. it gets to $q=8$, the inaccuracy of the fit of
the modulus start becoming relevant. For example, we did preliminary investigations
on a nonpublic $(8,+0.85,+0.85)$  and $(8,+0.80,0)$ waveform data computed using the BAM code
and presented in Ref.~\cite{Husa:2015iqa}, which indicate that this effect shows up
in this case, with fractional amplitude differences that grow up to $5\%$ in the first
$20M_{\rm BH}$ after merger. A detailed analysis of the performance of the template on these
datasets and, especially, ways to improve it, will be discussed in future work. For the
moment, we just warn the reader that our analytical post-merger template waveform
(either the primary fit or the interpolating one) may develop non-negligible inaccuracies
for large mass ratios (say $q\gtrsim 8$) and large spins (say  $\chi\gtrsim |0.8|$).
By contrast, we will show below that the template is certainly rather faithful
up to $q=3$ and spins up to $\pm 0.85$. A modified primary fitting ansatz that
(i) includes more parametric flexibility for the amplitude  and (ii) allows for an
effective representation of the oscillations entailed by the presence of mode mixings 
will be eventually necessary to improve the accuracy of the post-merger analytical 
template all over the parameter space~\cite{Nagar:2016prep}.

\section{Template waveform accuracy}
\label{sec:accuracy}
We assessed the accuracy of the primary fitting and interpolating procedures by 
cross-validating our template on a complementary SXS dataset, see
Table~\ref{tab:NRtest}. For the parameters corresponding to each of the validation 
waveforms, we constructed the analytic post-merger waveform using the coefficients 
in Table~\ref{tab:fitting} and computed phase and amplitude differences 
with the SXS waveform. Note however that the fits are used {\it only}
to compute $\bar{h}(\tau)$. By contrast, $\sigma_{1}$ is obtained, 
as above, by interpolating the QNMs data of E.~Berti~\cite{berti:qnms} 
on the final state  $(M_{\rm BH},J_{\rm BH})$ provided by the {\tt metadata.txt} 
SXS file\footnote{In principle one could have computed $\sigma_{1}$ using 
the fit for $\alpha_{1}$ of Table~\ref{tab:fitting} and computing the imaginary 
part as $\omega_{1}=\Delta\omega + M_{\rm BH}\omega_{22}^{\rm mrg}$ 
where also $\Delta\omega$ and $\omega_{22}^{\rm mrg}$ are provided by 
the fits of Table~\ref{tab:fitting}.  However, in doing so the combined inaccuracies 
of the two fits can make the computation of  $\omega_{1}$ rather inaccurate 
(up to $~10\%$) depending on the particular dataset. This approach cannot then
be followed, but we postpone to future analysis the construction of
a more accurate global interpolating fit for $\omega_{1}$.} .

\begin{figure}[t]
\center
\includegraphics[width=0.42\textwidth]{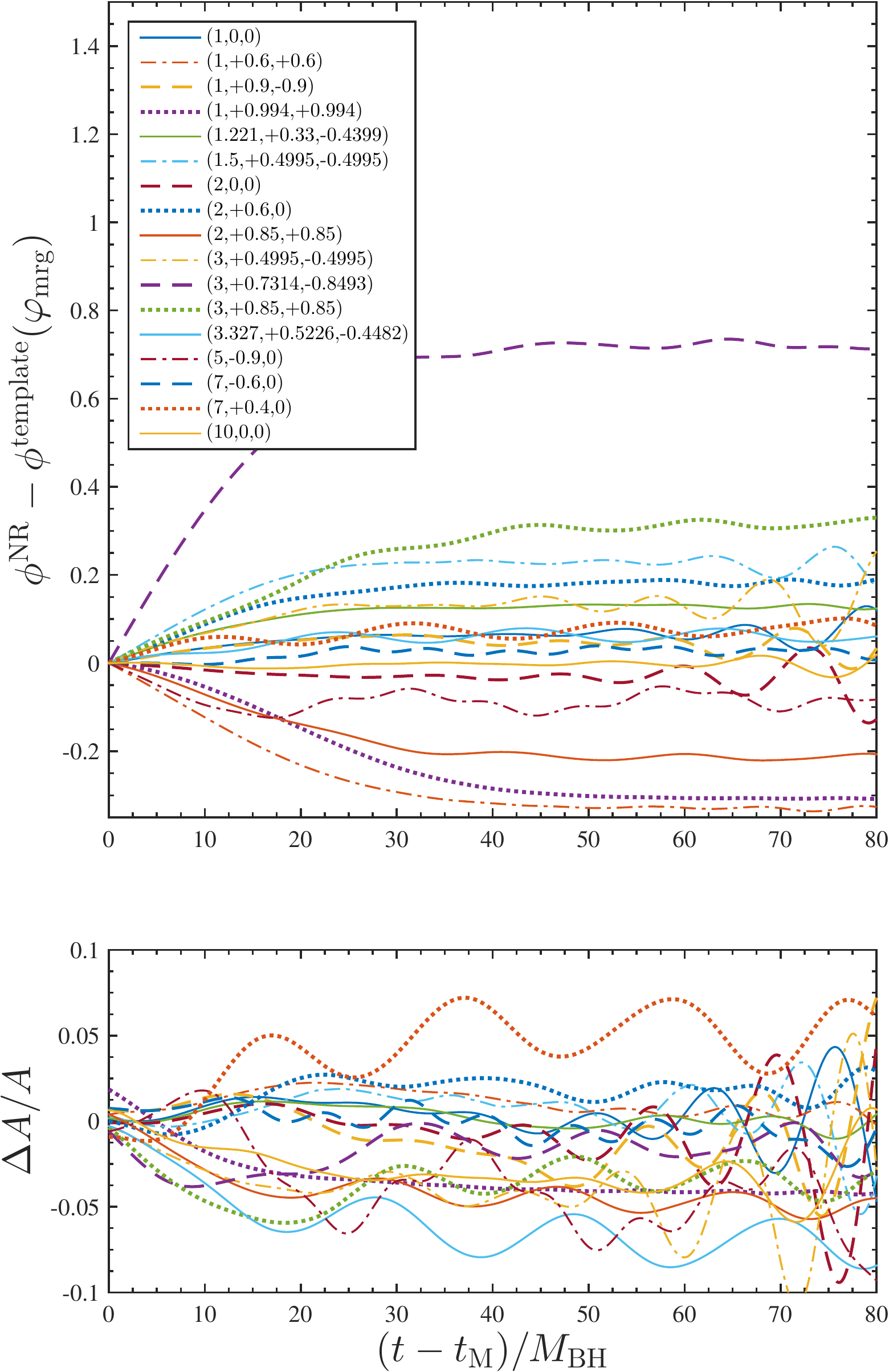}
\caption{\label{fig:alignatmrg} Straightforward evaluation of the
 performance of the general post-merger template obtained from
 Eq.~\eqref{eq:Yvsa0} and Table~\ref{tab:fitting}. The two waveforms
 are aligned by imposing the phase difference is zero at
 merger point. The corresponding NR phases at merger (that is subtracted
 from the analytic one) are listed in Table~\ref{tab:NRtest}.}
\end{figure}

\begin{figure}[t]
\center
\includegraphics[width=0.42\textwidth]{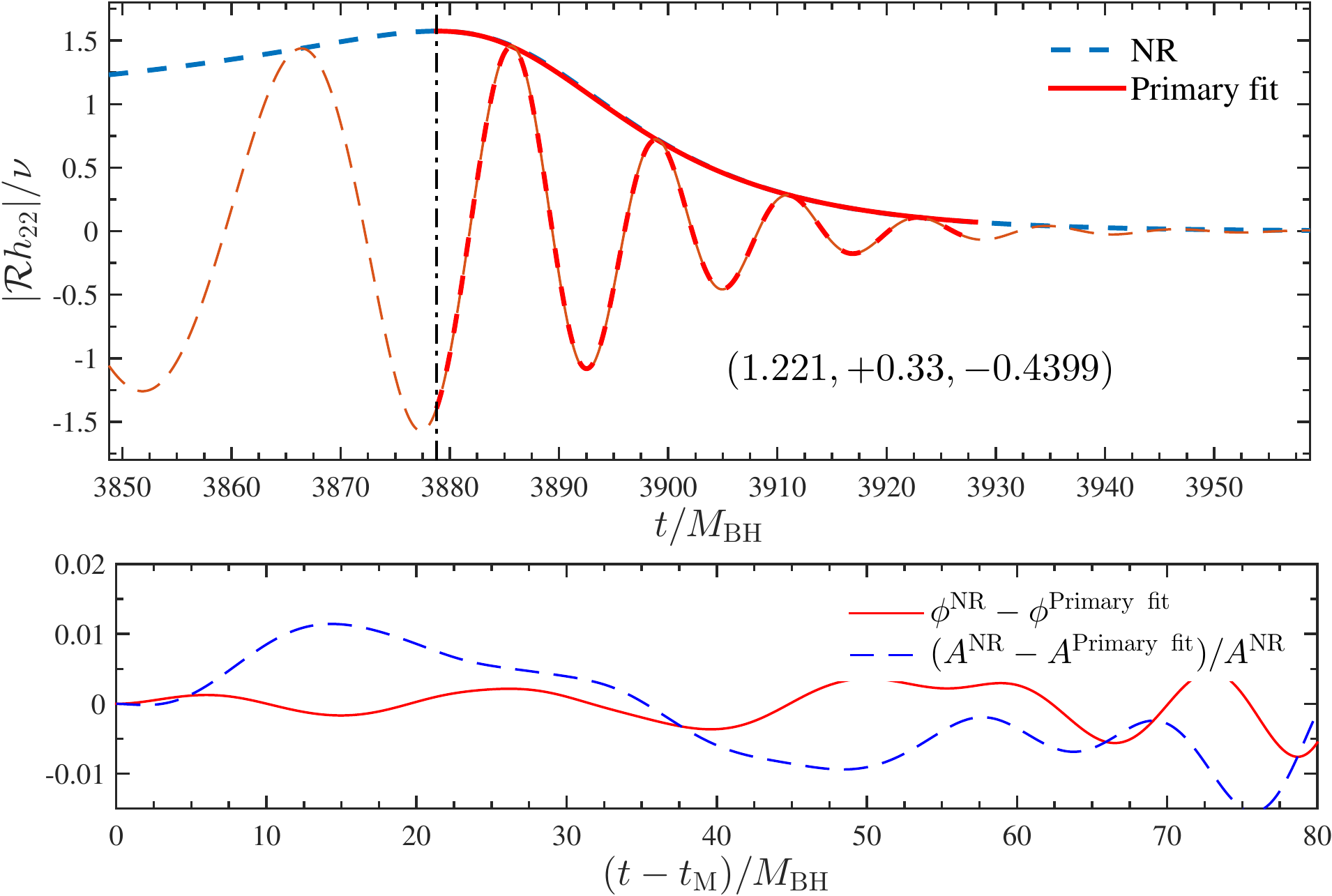}
\caption{\label{fig:GW150914}Performance of the primary fit on dataset SXS:BBH:0305.
The thick red line in the top panel marks the time interval where the fit 
is actually done. The phase difference at merger is consistent with the general 
interpolating fit after time and phase alignment (green line in Fig.~\ref{fig:figFit}).}
\end{figure}

The performance of the interpolated analytical waveform against the
numerical one is evaluated by means of two kind of phasing comparisons. 
First, the two waveforms are aligned just in {\it phase}, imposing that the phase 
difference is 0 at the moment of merger. This comparison aims at 
providing a precise idea of the accuracy of the interpolated fit with respect to the primary fits.
The result is presented in Fig.~\ref{fig:alignatmrg}. The worse performance corresponds
to SXS:BBH:0292, with $(3,0.7314,-0.8493)$, a dataset  not used for the template construction, 
where the phase difference grows up to 0.7 rad over the first $30M_{\rm BH}$ after merger. 
This figure illustrates the intrinsic limitations of our post merger interpolating fit,
that are mostly due to the limited amount of NR waveform data that were available
when this work was started\footnote{As mentioned above, when this paper was under review,
  on October 31st 2016, the SXS collaboration made public another 95 spin-aligned
  waveforms with mass ratio $q$ varying between 1 and 3, originally
  presented in~\cite{Chu:2015kft,Kumar:2016dhh}. This data are not included in
  the template construction, though a few of them are used to validate the
  interpolation outside its ``calibration'' domain. The new datasets used to this
  aim are: SXS:BBH:0257, SXS:BBH:0211, SXS:BBH:0292, SXS:BBH:0293. The incorporation of,
at least part of, this large amount of NR data in the template construction, together
with a few structural modifications outlined above, is expected to strongly improve
its performance, and will be discussed elsewhere.}.
Such large phase differences may be relevant when the interpolating fit is used
to provide the post merger waveform in EOB models, as the one of Refs.~\cite{Nagar:2015xqa}
and more recently of Ref.~\cite{Bohe:2016gbl}, whose ringdown is calibrated to a much larger
sets of NR SXS waveforms that also include those of Refs.~\cite{Chu:2015kft,Kumar:2016dhh}
publicly released on October~31st 2016. The precise evaluation of the quality of the
current post merger model for EOB purposes is outside the scope of this work and will
also be analyzed elsewhere. Note, however, that the quality of the primary fitting
procedure for a single NR dataset is on average rather good; it is illustrated in
Fig.~\ref{fig:GW150914}, for the case of SXS:BBH:0305. For this GW150914-like waveform,
the phase difference is of the order of 0.01~rad and the amplitude (relative)
difference of about $1\%$. We will see below that this behavior is essentially typical
fro most the of the NR dataset analyzed, with a few exceptions that we discuss explicitly.

When the analytic post merger waveform is used as a template for parameter
estimation, it is actually defined modulo an arbitrary time and phase shift.
As a consequence, it also makes sense to compare the analytical and numerical
waveform by aligning them fixing these two arbitrary constants.
We use here the alignment procedure introduced in Sec.~VA of Ref. ~\cite{Baiotti:2011am}
and extensively used in subsequent EOB/NR works (see e.g. ~\cite{Nagar:2015xqa} and
references therein). The phase and time shift are chosen so that the phase
difference is minimized over a small frequency interval after merger. We use an
interval because, in general, in this way the alignment procedure is more
robust and less affected by numerical artefacts that may be present
in the numerical waveforms. The minimization interval is chosen to be 
$M_{\rm BH}[\omega_{\rm L},\omega_{\rm R}]=M_{\rm BH}\omega_{\rm mrg}[1.05,1.20]$,
that ends always well before the final fundamental QNM frequency is reached.

\begin{figure}[t]
\center
\includegraphics[width=0.42\textwidth]{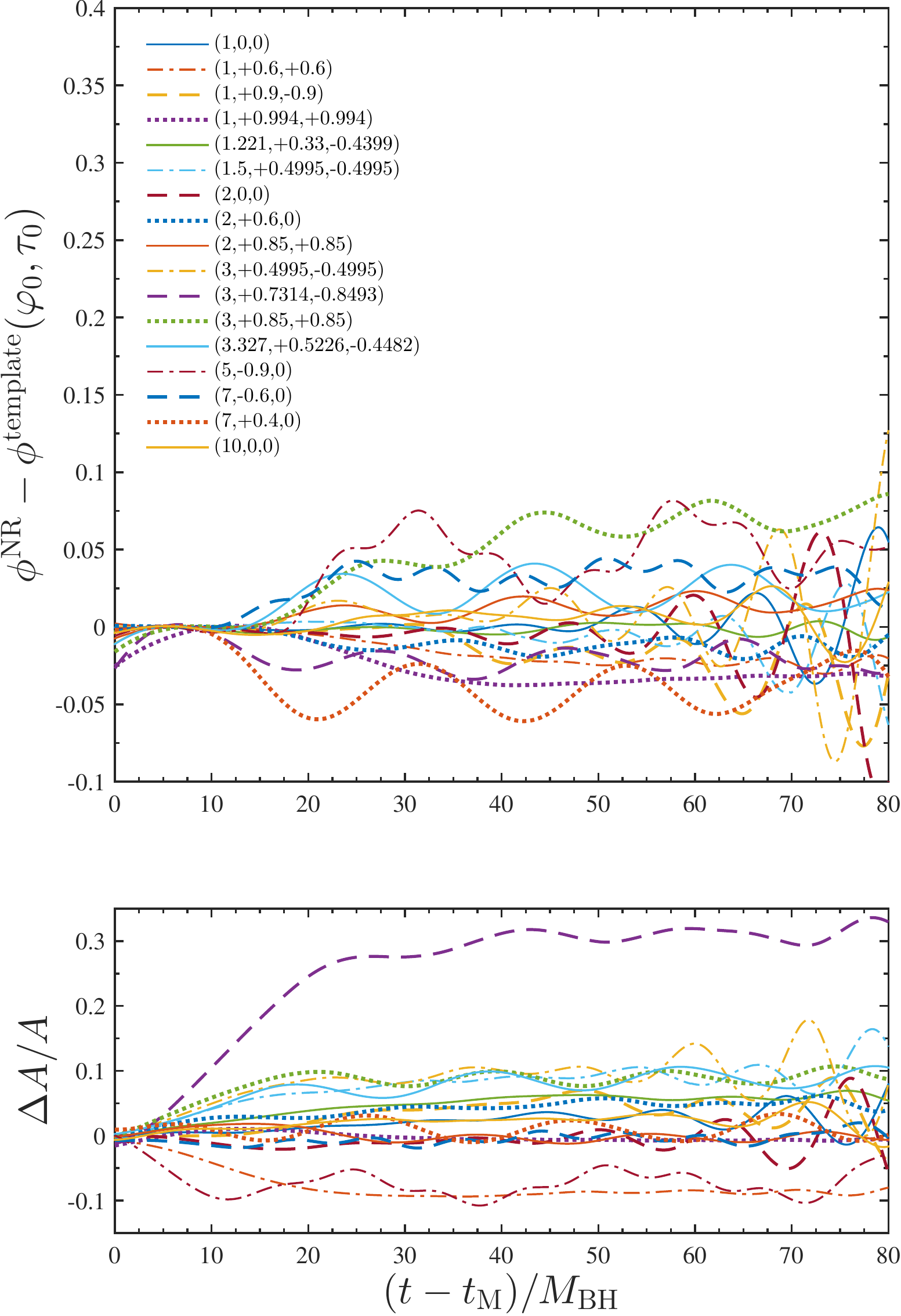}
\caption{\label{fig:figFit}Performance of the general post-merger template 
obtained from Eq.~\eqref{eq:Yvsa0} with the coefficients given in 
Table~\ref{tab:fitting}. For each dataset, the analytical template
is aligned to the corresponding NR waveform by fixing an arbitrary
phase and time shift.The phase difference is usually compatible with 
the typical accuracy of the primary fit (see Fig.~\ref{fig:GW150914}),
except for some dataset with large mass ratio and high spins.}
\end{figure}
The result of the time and phase shift is illustrated in Fig.~\ref{fig:figFit}.
The phase difference (top panel) is, in general, oscillating around zero
for most datasets, with the largest values of the oscillations, $\sim 0.1$~rads,
arising for late times, where the corresponding NR waveforms get progressively
dominated by numerical oscillations (e.g., due to the radius extrapolation
procedure, see also discussion in~\cite{Damour:2014yha}). Note that, however,
this {\it is not} the case for few datasets (e.g., $(5,-0.9,0)$) where the
phase difference does not average zero even after the alignment procedure,
showing then qualitative differences with respect to the straightforward alignment
at merger. This is probably due to the lack of a proper
representation of the interference between prograde and retrograde QNMs for this
particular dataset; we will see below that this behavior is mirrored in systematics
in the determination of the parameters. By contrast, the {\it fractional}
amplitude differences (bottom panel) tend to be $5\% \leq \Delta A/A \leq 10\%$, 
with similar increasing oscillations as time grows. 
Illustrated in the figures is the result of the cumulative effect of two sources of
uncertainty: (i) the intrinsic limitation of the ansatz for the primary fit, 
Eqs.~\eqref{eq:barA}-\eqref{eq:barPhi}; (ii) the fact that the general interpolation
all over the parameter space is done starting from a rather small number of 
training datasets, most of which are equal-mass, spin-aligned configurations.
We will not discuss point (ii), since a thorough analysis would require to
redo our analysis including progressively more of the datasets 
of Ref.~\cite{Chu:2015kft} recently included in the SXS catalog, but rather
just focus on point (i). For most of the BBHs configurations considered in
this paper, the primary fit performs similarly to the
GW150915-like dataset depicted in Fig.~\ref{fig:GW150914}, with phase
differences (computed just doing the straight phase-alignment at merger
point) oscillating between $\pm 0.005$~rad (at maximum) all over the 
post merger phase and fractional amplitude differences  $\Delta A/A$ 
of order $1\%$ or smaller.  This is typical for the equal-mass, equal-spin datasets, while
it worsens when {\it both} the mass ratio and the spin 
increase. For example, for $(3,+0.85,+0.85)$, that is not used for the 
construction of the interpolating fit, the phase difference yielded by the primary
fit oscillates between $-0.01$~and~$+0.015$~rads; things get even worse
for $(5,-0.90,0)$ (that is actually part of the template construction), 
where the phase differences is found to oscillate between  $\pm 0.04$~rad 
across the full post-merger phase. These phase difference are actually
{\it rather large} in this context and will propagate (and possibly increase)
in the construction of the interpolating template. Inspecting Fig.~\ref{fig:figFit}
one sees that  the phase difference for $(5,-0.90,0)$ starts with zero, but then
increases and oscillates around 0.05~rad all over the post-merger phase.
In general, to improve our analytical model further in order to have it
more reliable in specific corners of the parameter space one would need 
(i) more NR simulations of asymmetric systems ($\nu\neq 1/4$, $\ta_1\neq \ta_2$, etc.,
see e.g.~\cite{Kumar:2016dhh,Jani:2016wkt,Husa:2015iqa}), possibly
taking into account also information coming from large-mass-ratio 
waveforms~\cite{Harms:2014dqa,Taracchini:2014zpa,Bernuzzi:2011aj},
and (ii) different ansatz for the primary fitting template,
Eqs.~\eqref{eq:barA}-\eqref{eq:barPhi}, so as to take into account
the mixing between the retrograde and prograde fundamental QNMs. 
In any case, despite the limitations outlined above, our global interpolation 
scheme still provides a complete, and rather reliable, description
of the {\it full} postmerger waveforms that explicitly depends
on $(m_1,m_2,\ta_1,\ta_2,M_{\rm BH},\sigma_1)$ 
(as well as an initial arbitrary phase $\phi_0$ and time $t_0$).
Such postmerger waveform could be therefore used, for instance,
to improve on the existing inspiral-merger-ringdown consistency
test presented in Ref.~\cite{TheLIGOScientific:2016src} as well
as on the measurement of the least-damped QNM parameters.

\section{Data analysis}
\label{sec:da_stuff}
\begin{figure*}[t]
\begin{center}
  \includegraphics[width=0.35\textwidth]{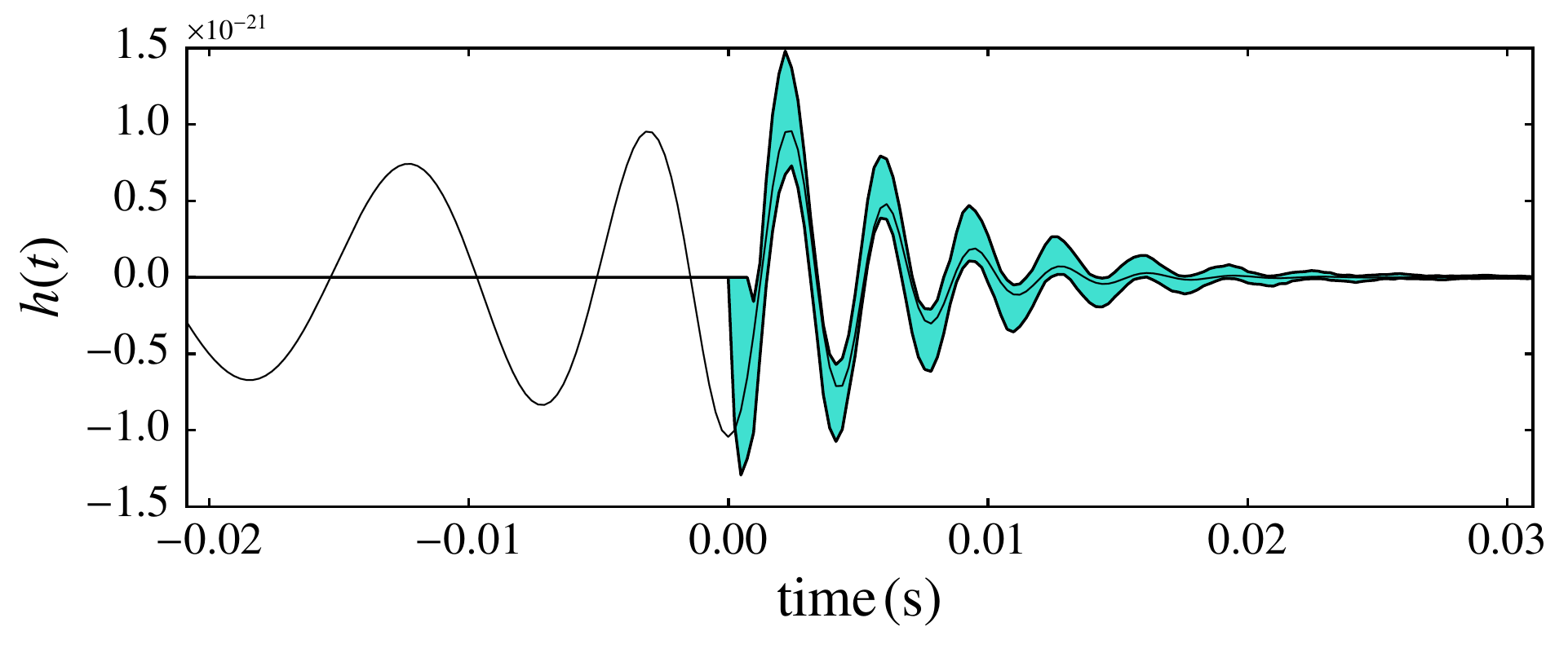}
  \hspace{10mm}
  \includegraphics[width=0.35\textwidth]{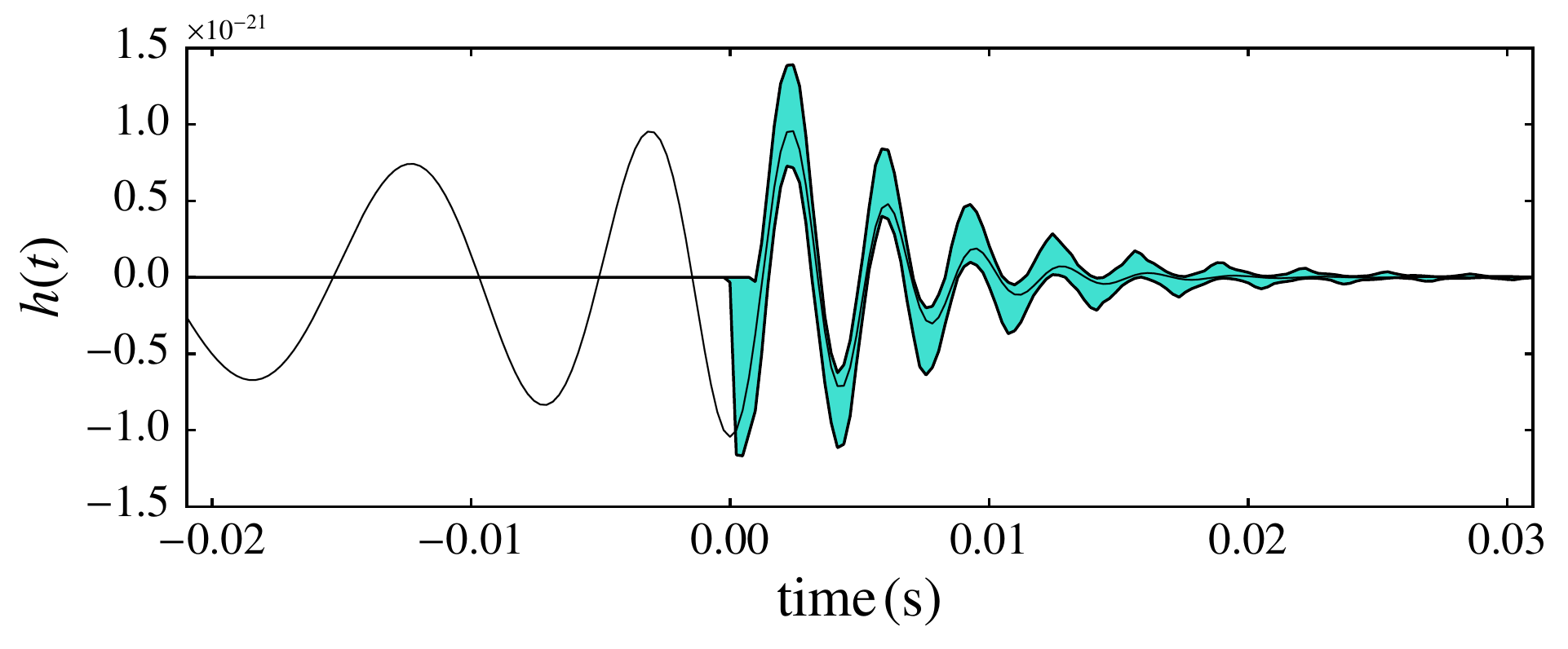}\\
  \includegraphics[width=0.35\textwidth]{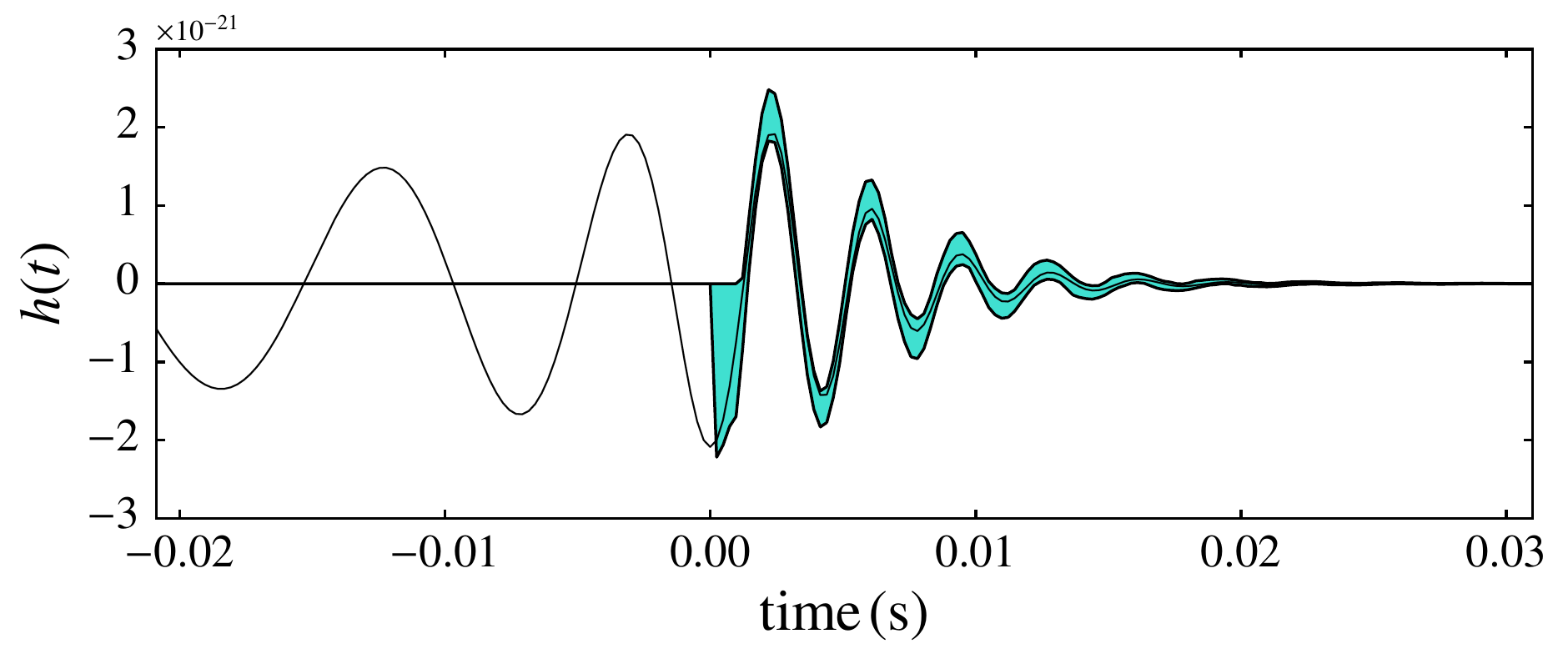}
  \hspace{10mm}
  \includegraphics[width=0.35\textwidth]{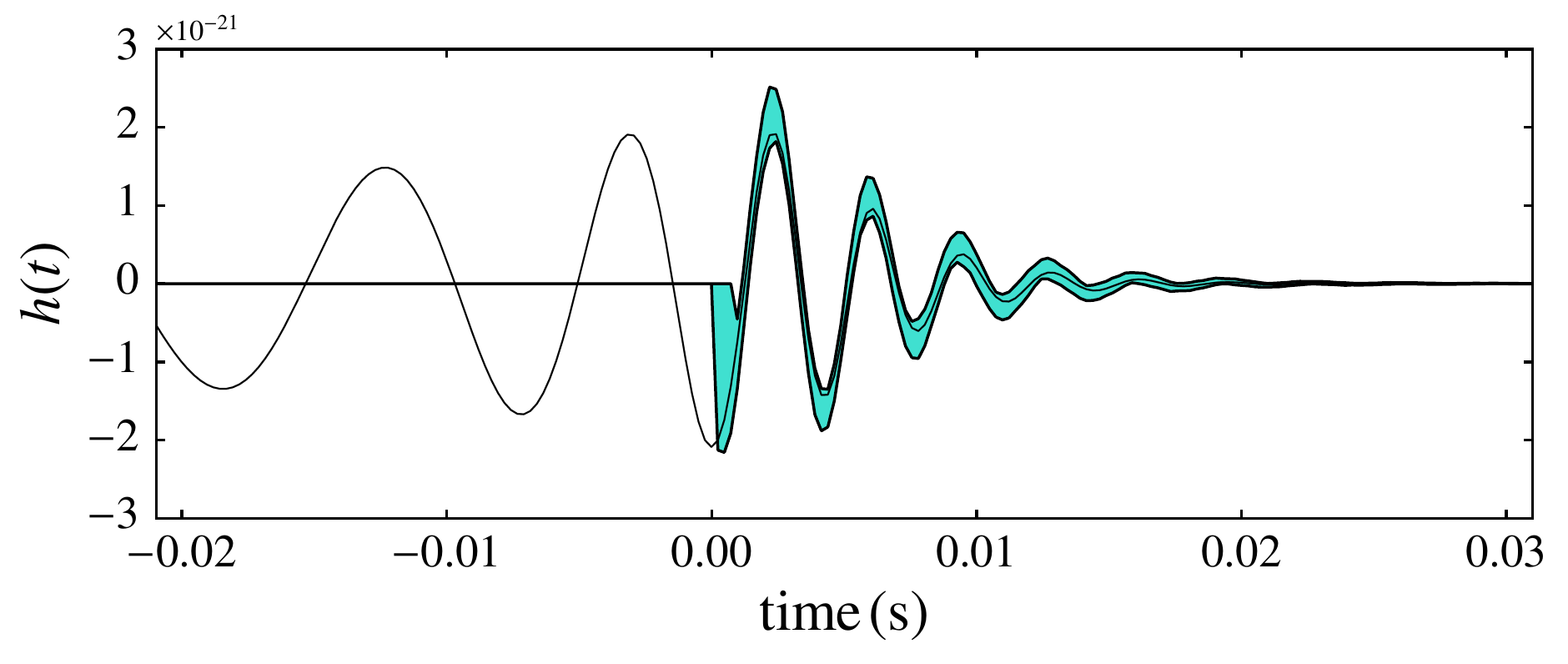}\\
  \includegraphics[width=0.35\textwidth]{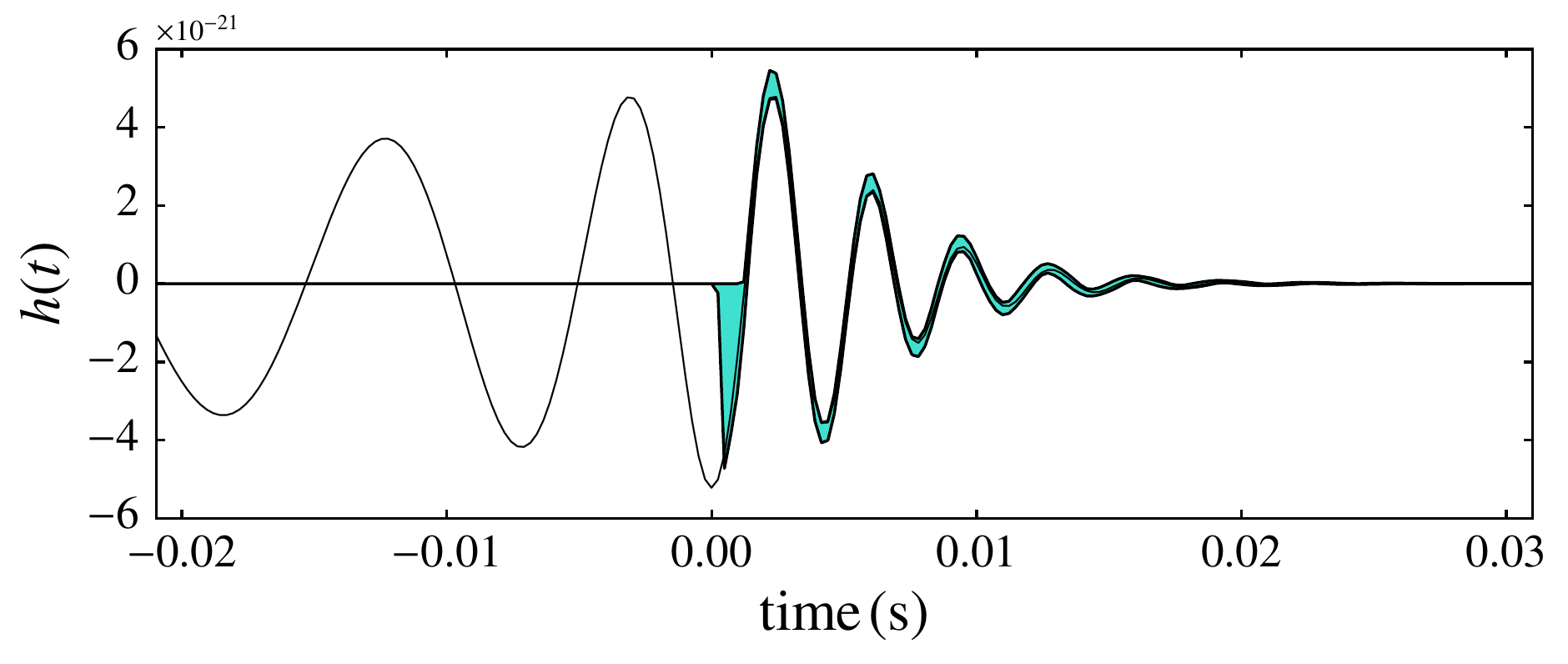}
  \hspace{10mm}
  \includegraphics[width=0.35\textwidth]{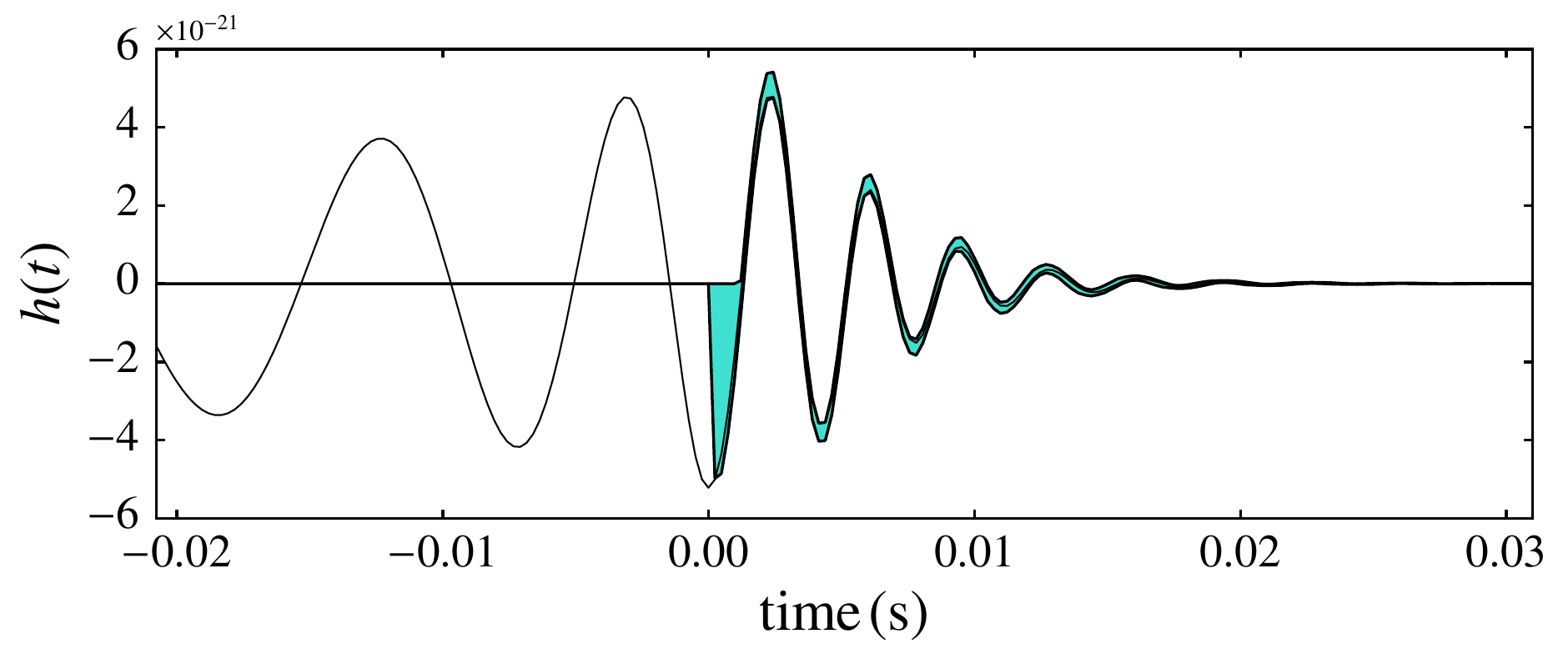}\\
  \includegraphics[width=0.35\textwidth]{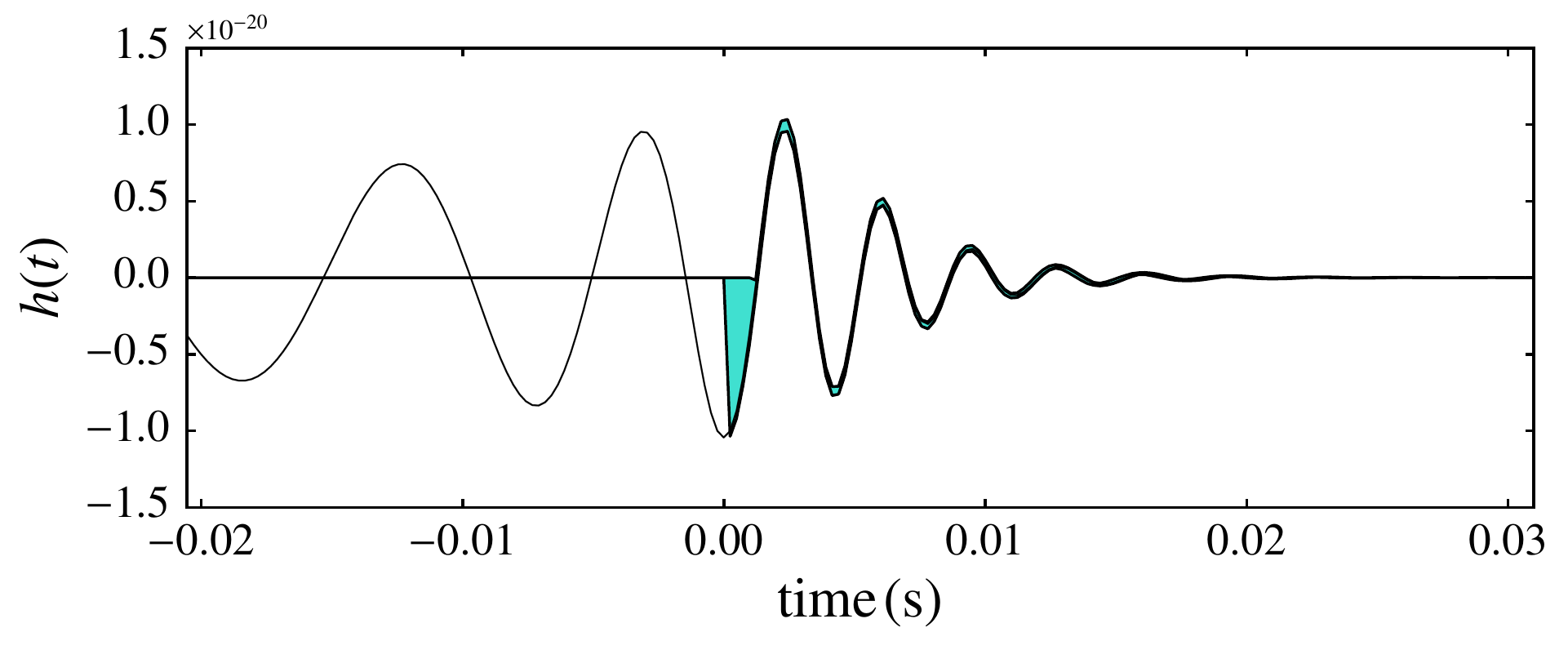}
  \hspace{10mm}
  \includegraphics[width=0.35\textwidth]{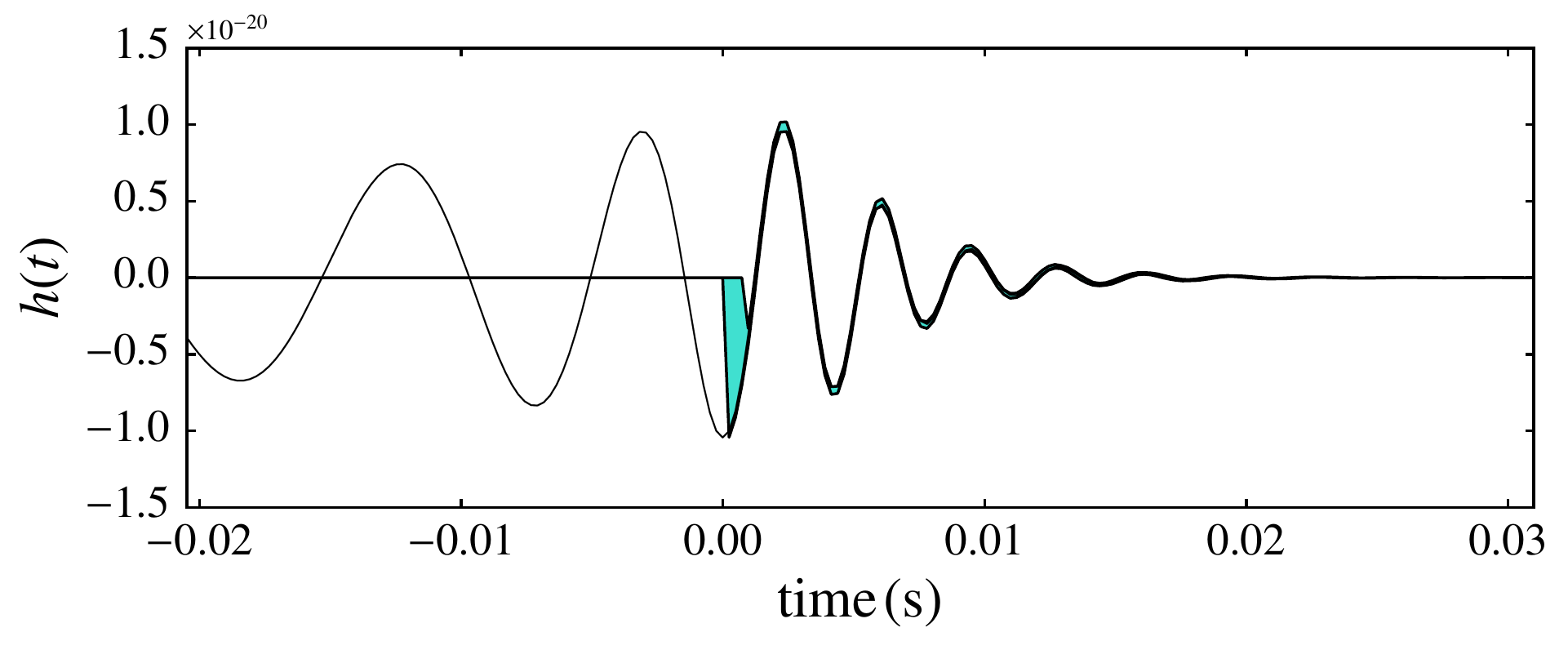}
  \caption{\label{fig:post_GE_wf}Left column: reconstructed post-merger
    waveform for the {\it generic} (GE) case (no relations among parameters are assumed) 
    and corresponding $90\%$ confidence region (shaded area) for the 
    GW150914-like dataset SXS:BBH:0305 with post-merger SNR = 10 (top panel), 20, 50 and 100 (bottom panel).
    Right column: the same, but for the {\it constrained} (CO) case, that assumes that an estimate for the component
    masses and spins exists. In all cases, the post-merger waveform is reconstructed very accurately, 
    with uncertainty decreasing as the post-merger SNR increases.}
\end{center}
\end{figure*}
\begin{figure*}[t]
\begin{center}
\begin{tabular}{cc}
  \includegraphics[width=0.32\textwidth]{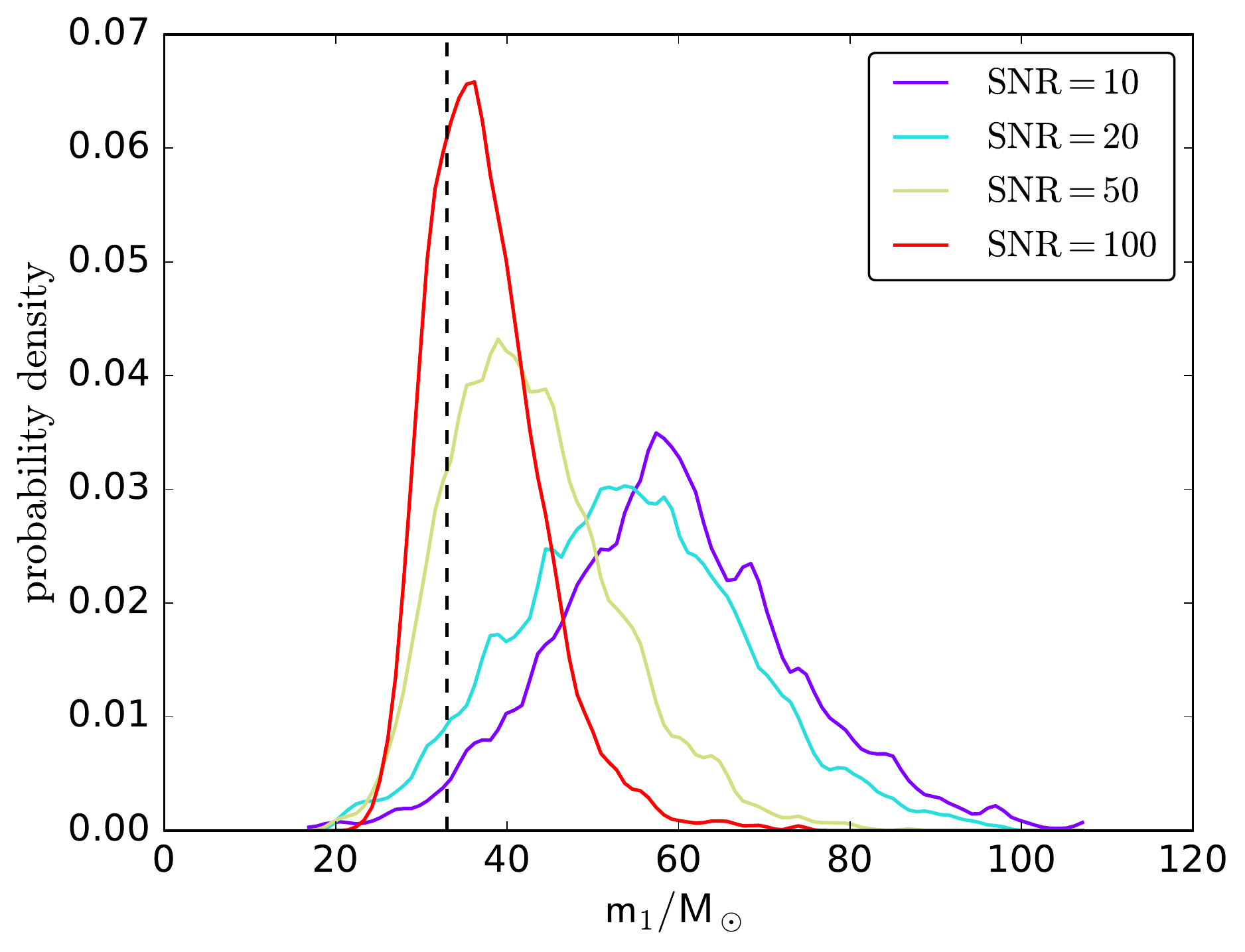}
  \hspace{10mm}
   \includegraphics[width=0.32\textwidth]{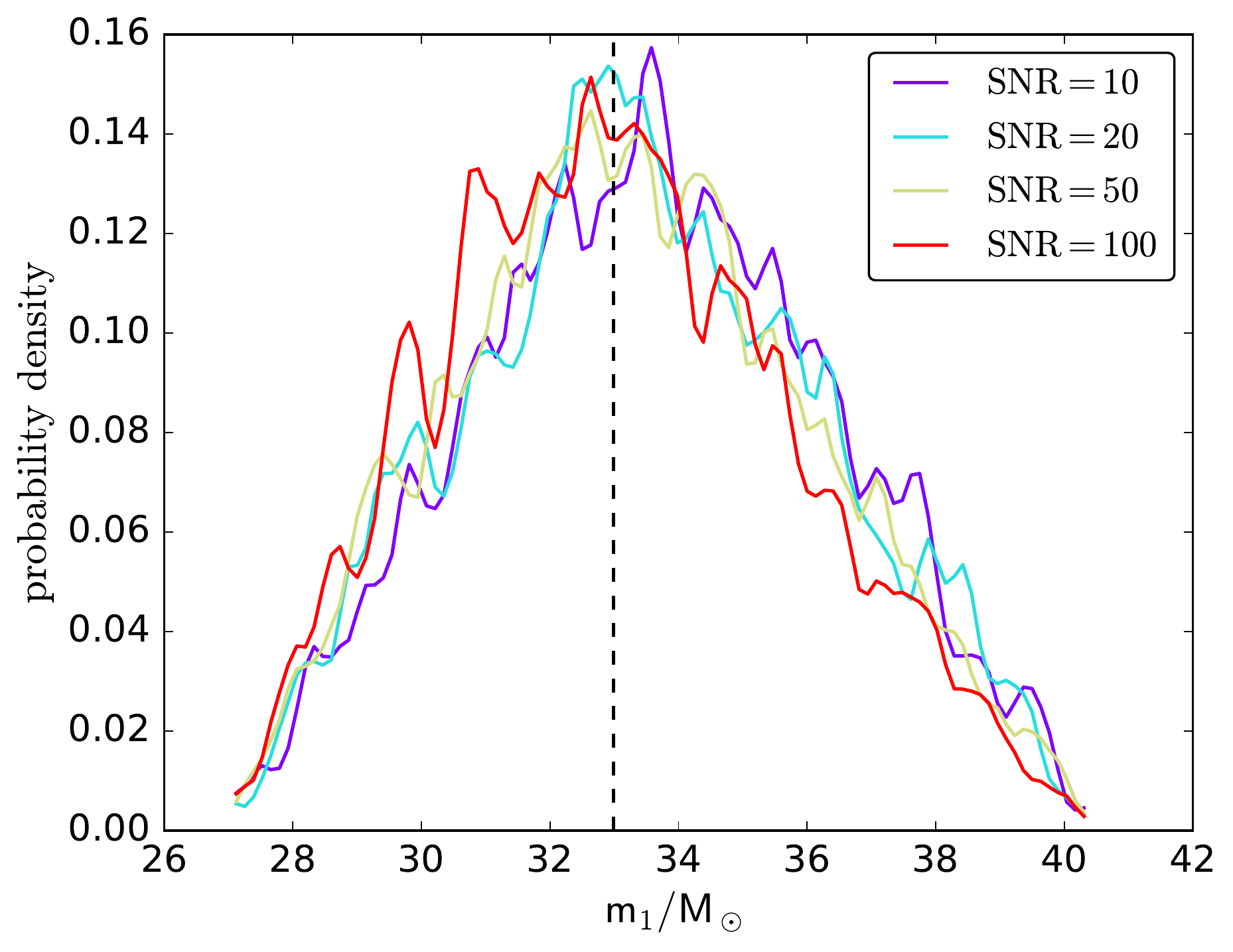}\\
  \includegraphics[width=0.32\textwidth]{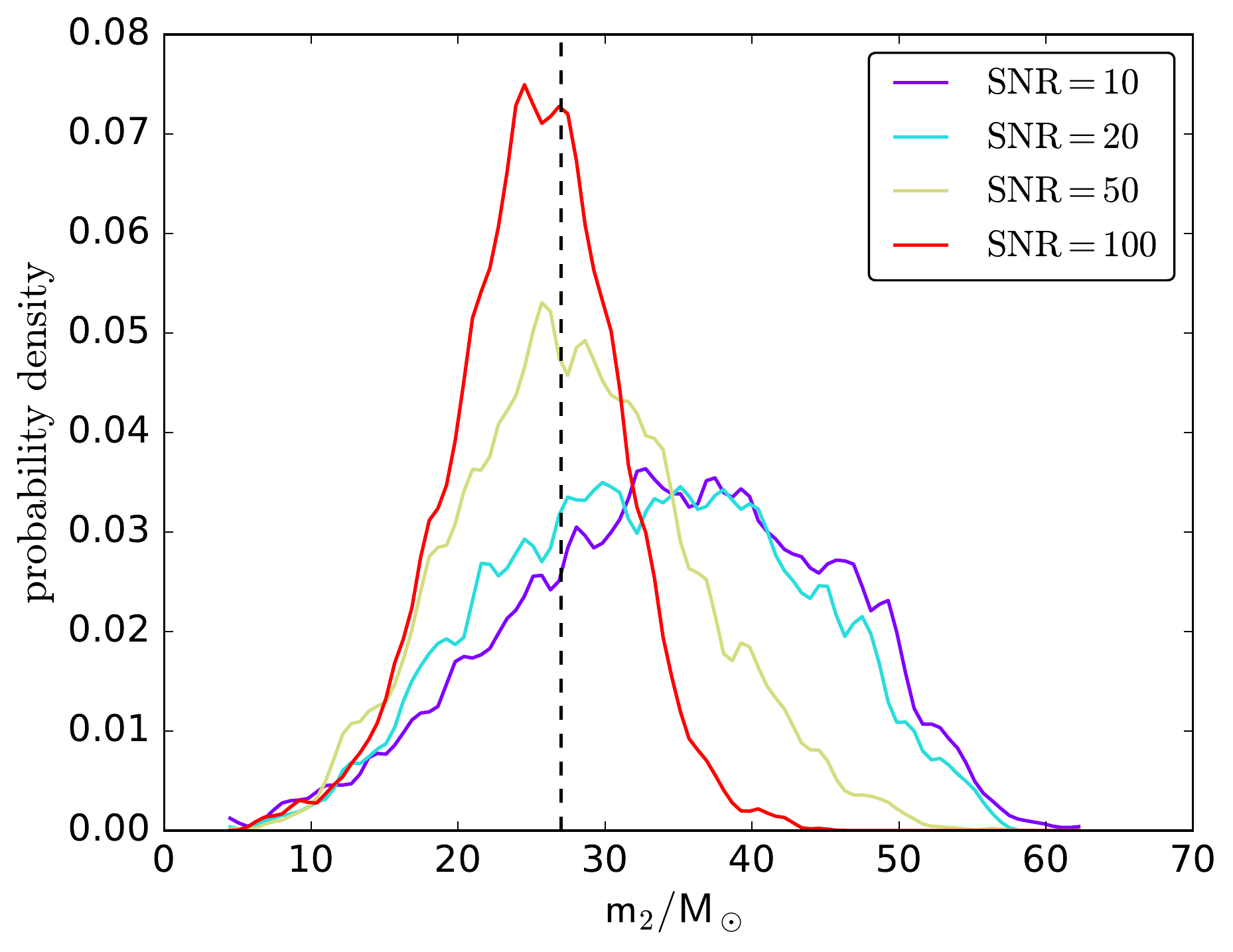}
    \hspace{10mm}
  \includegraphics[width=0.32\textwidth]{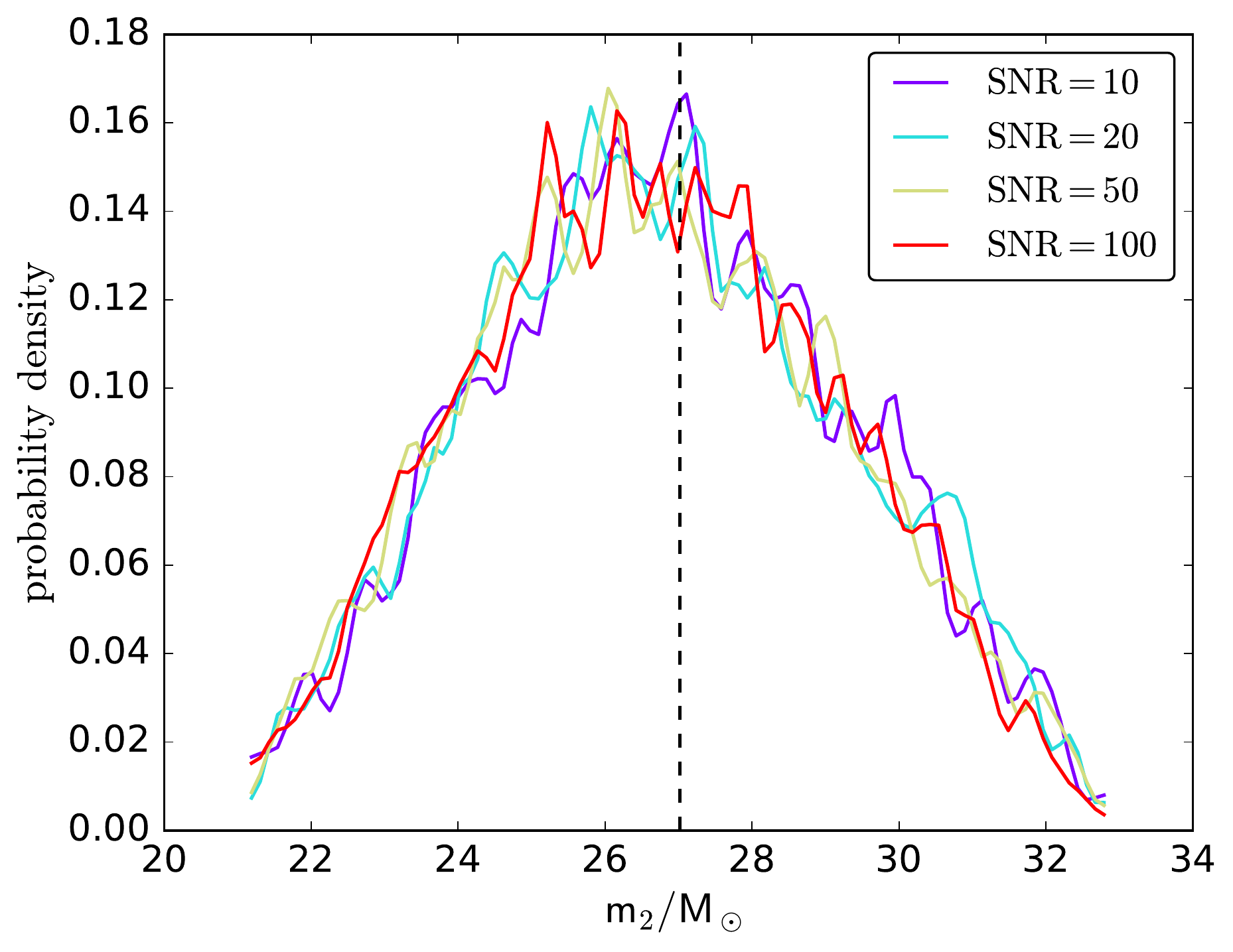}\\
  \includegraphics[width=0.32\textwidth]{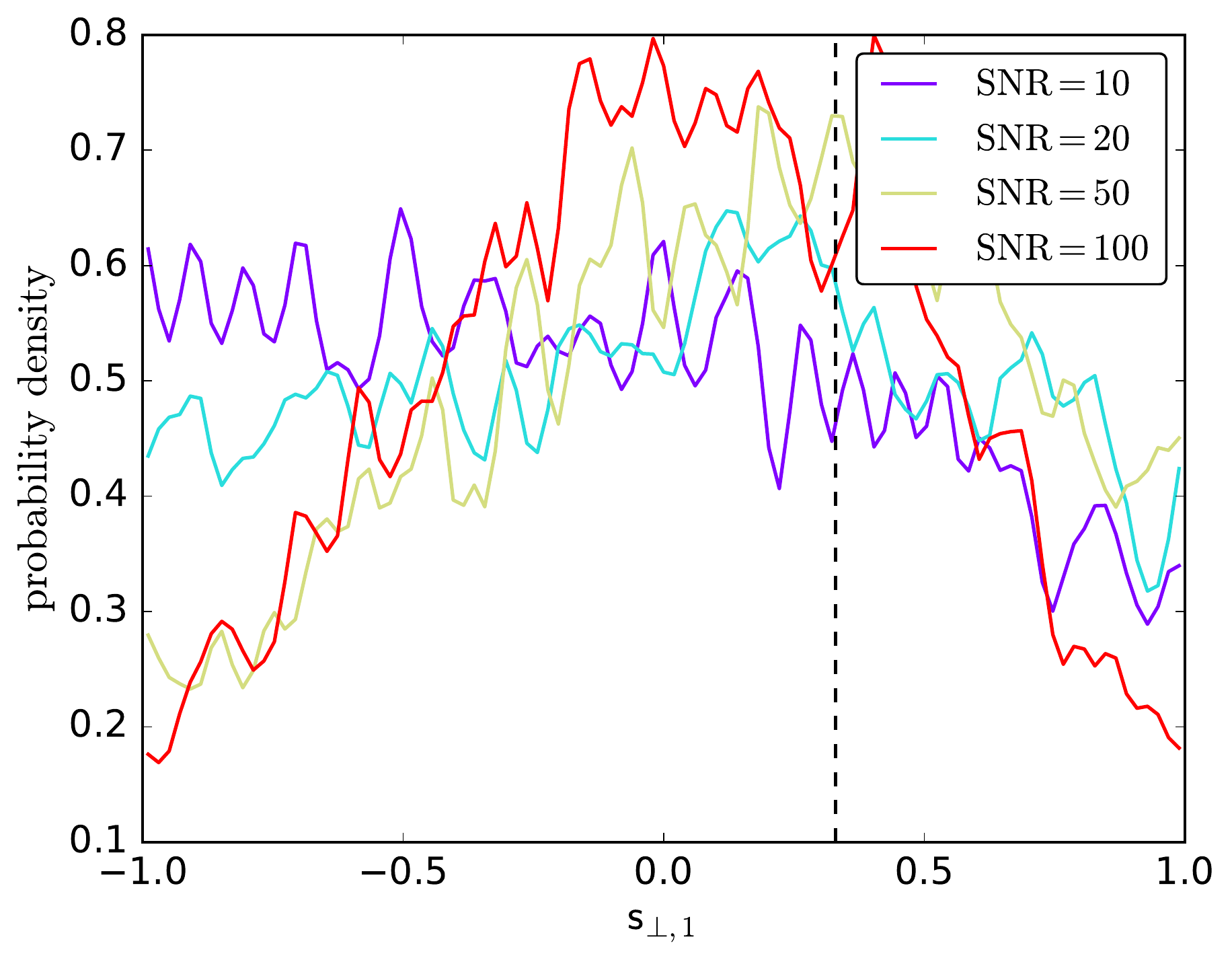}
  \hspace{10mm}
  \includegraphics[width=0.32\textwidth]{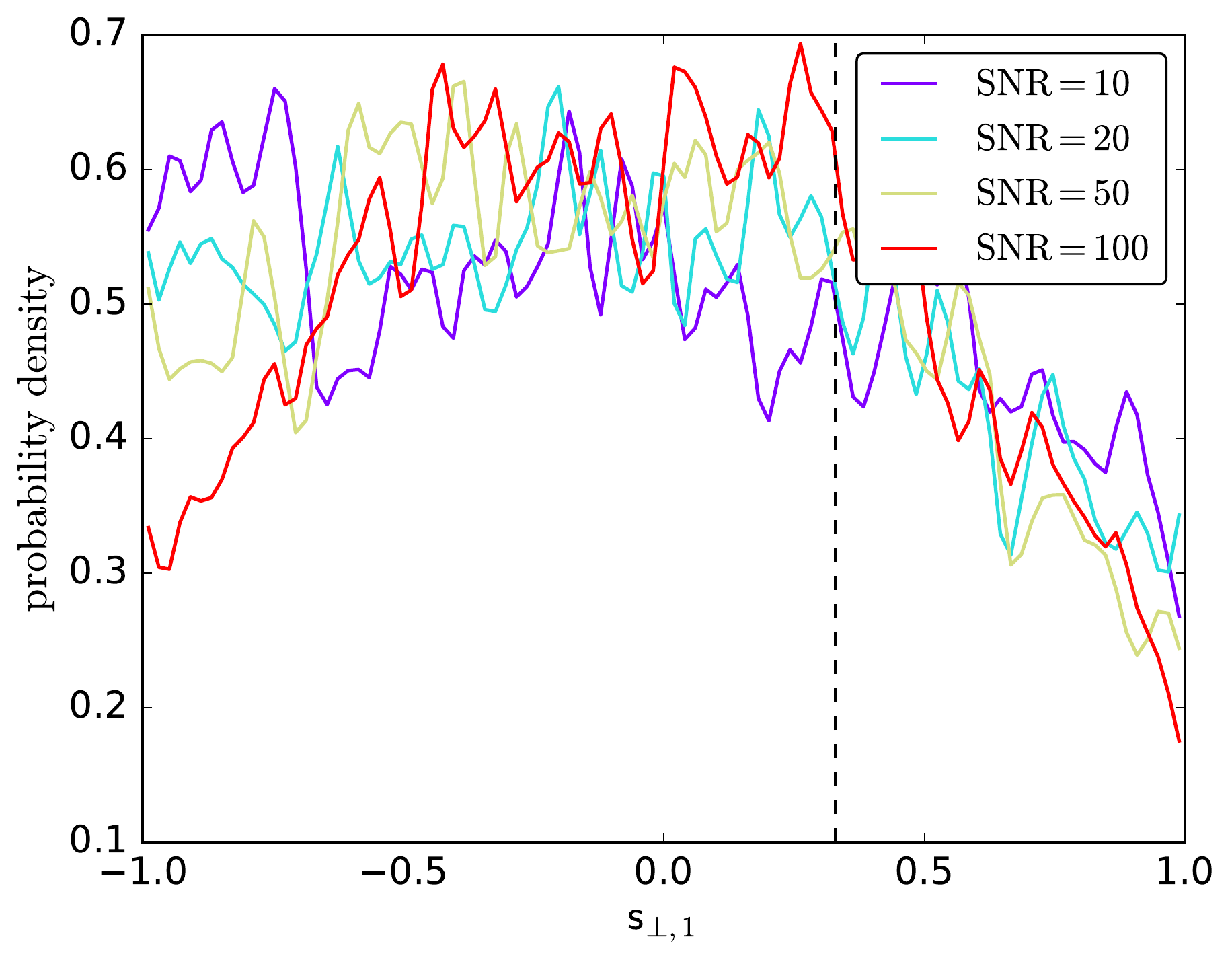}\\
  \includegraphics[width=0.32\textwidth]{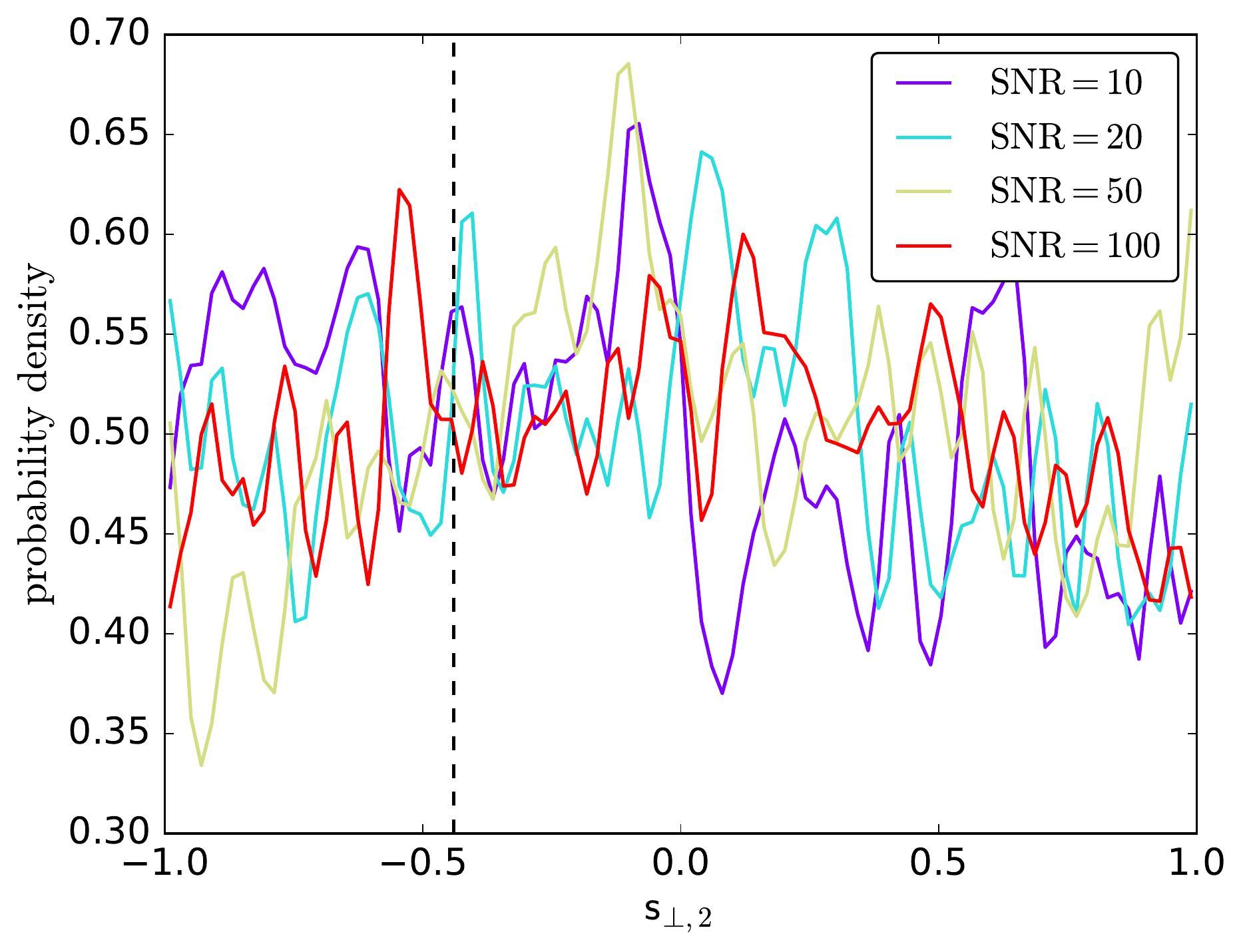}
  \hspace{10mm}
  \includegraphics[width=0.32\textwidth]{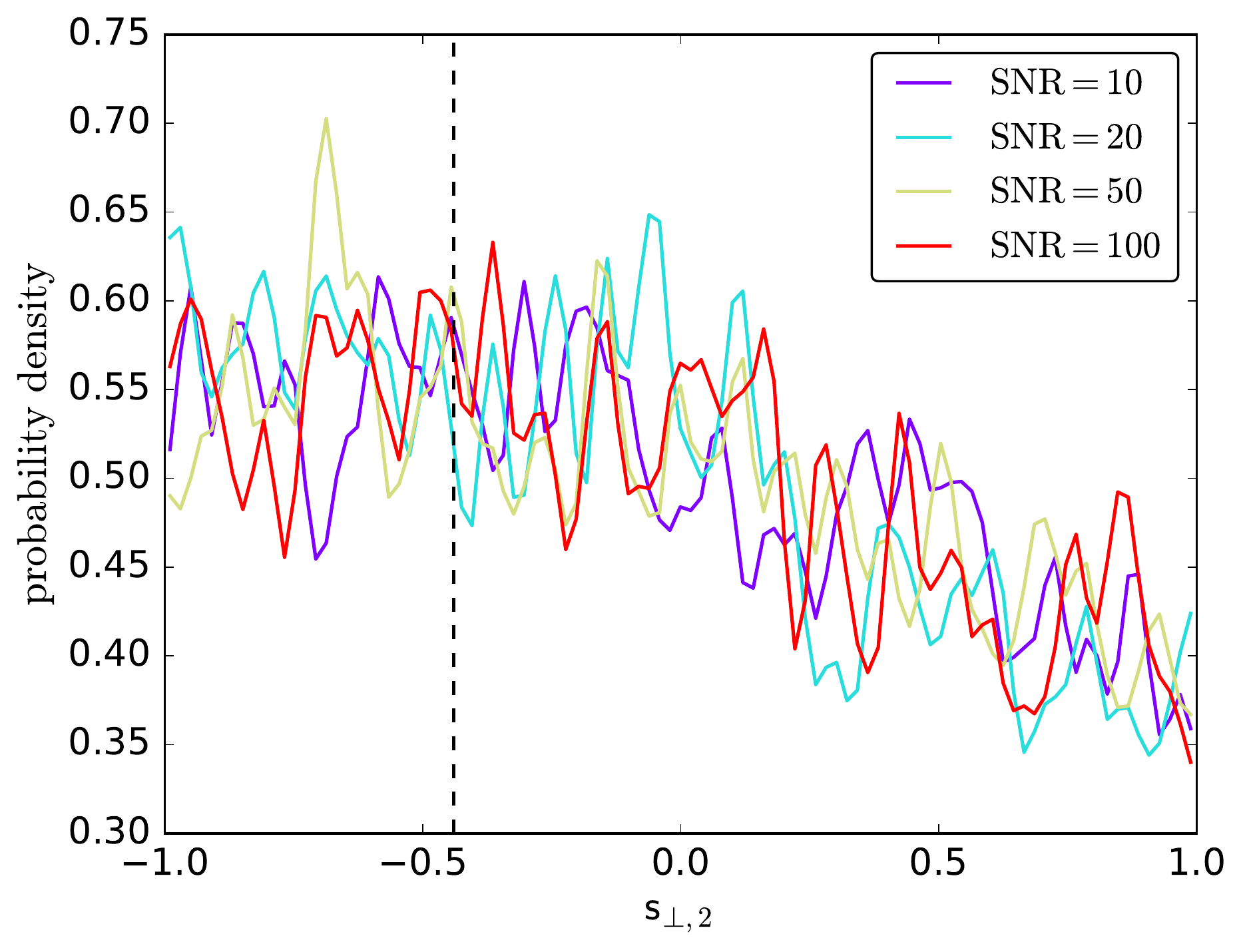}\\
\end{tabular}
\caption{\label{fig:post_GE1} Measurability of the component masses $(m_{1},m_{2})$ and 
dimensionless spins $(s_{\bot,1},s_{\bot,2})$ for the GW150914-like dataset SXS:BBH:0305.
The left panels show posterior distributions probabilities for the {\it generic} (GE)
 case (no relations among parameters are assumed) with post merger SNR = 10, 20, 50 and 100.  
 The right panels refer to the {\it constrained} (CO) case, that assumes that an 
  estimate for the component masses and spins exists. The vertical line in each panel 
  indicates the correct value of the measured parameter.}
\end{center}
\end{figure*}
\begin{figure*}[t]
\begin{center}
\begin{tabular}{cc}
  \includegraphics[width=0.32\textwidth]{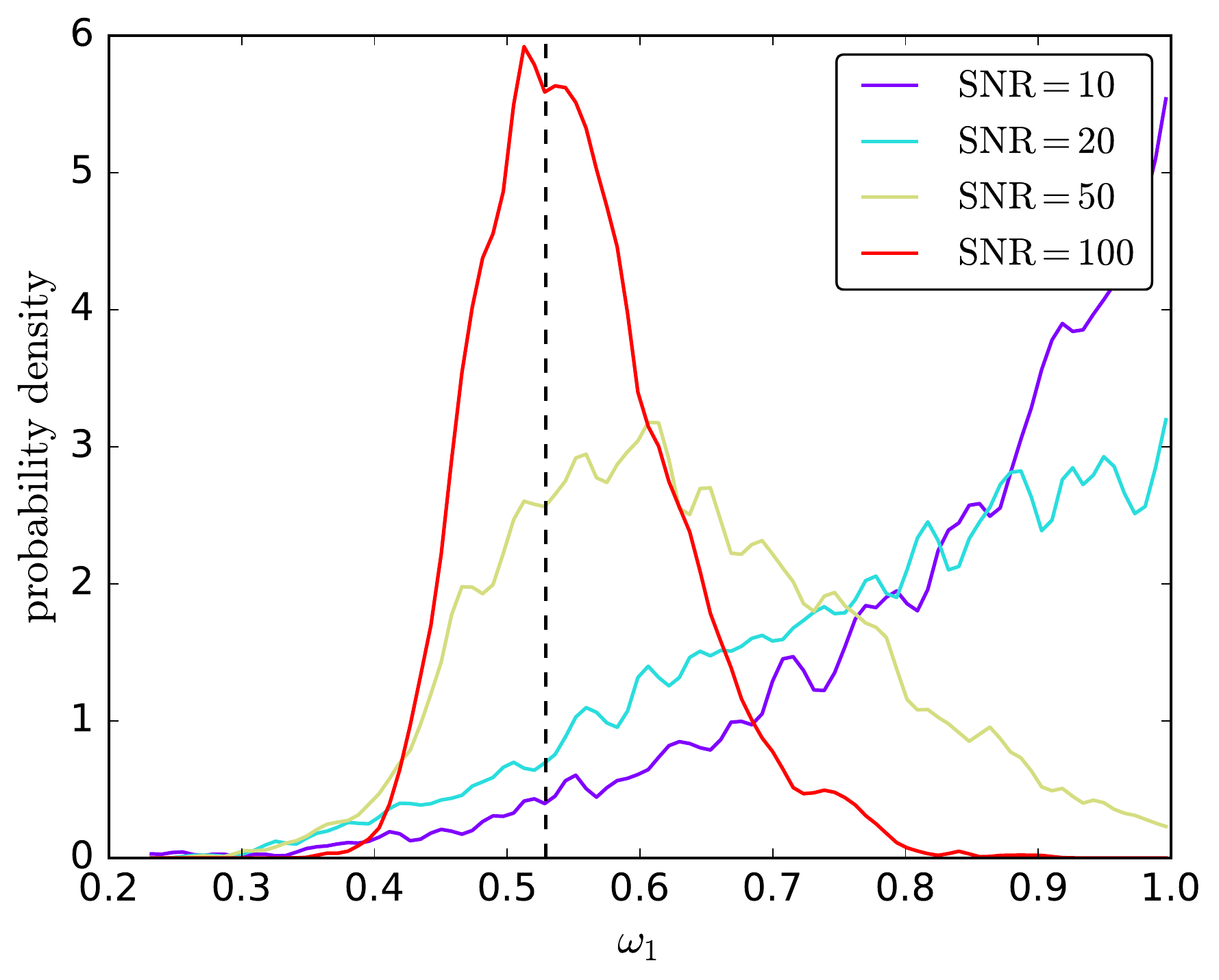}
    \hspace{10mm}
  \includegraphics[width=0.32\textwidth]{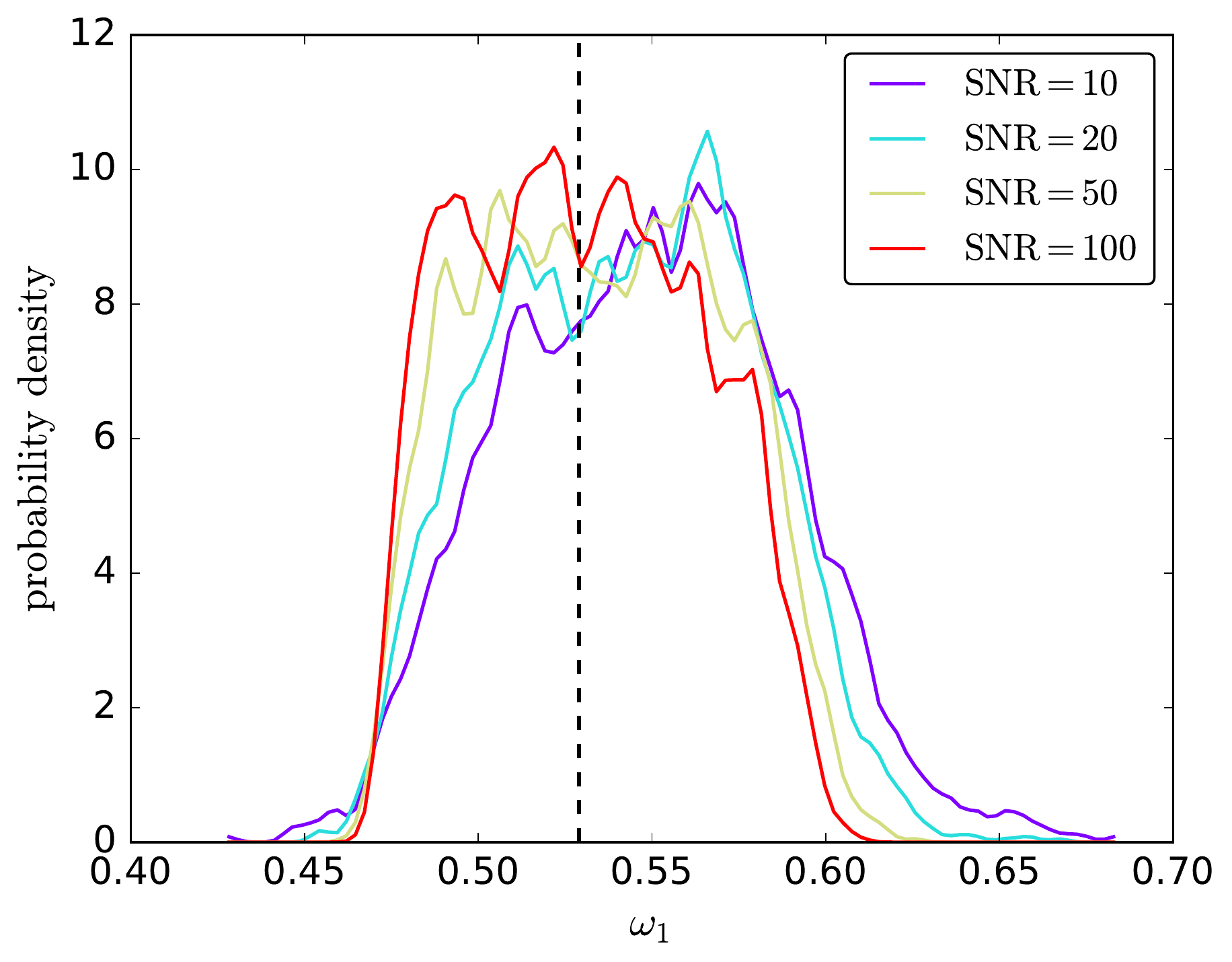}\\
  \includegraphics[width=0.32\textwidth]{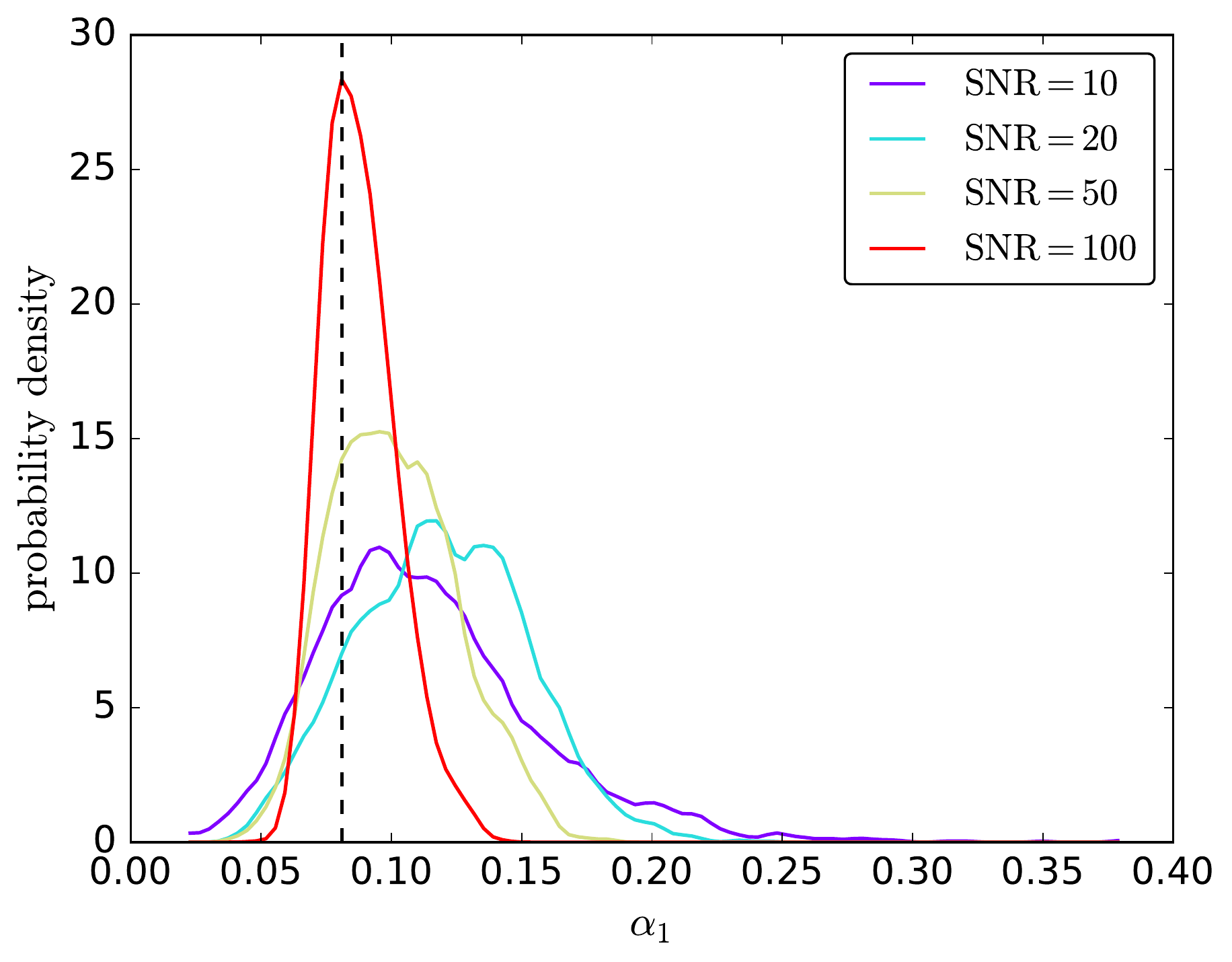}
    \hspace{10mm}
  \includegraphics[width=0.32\textwidth]{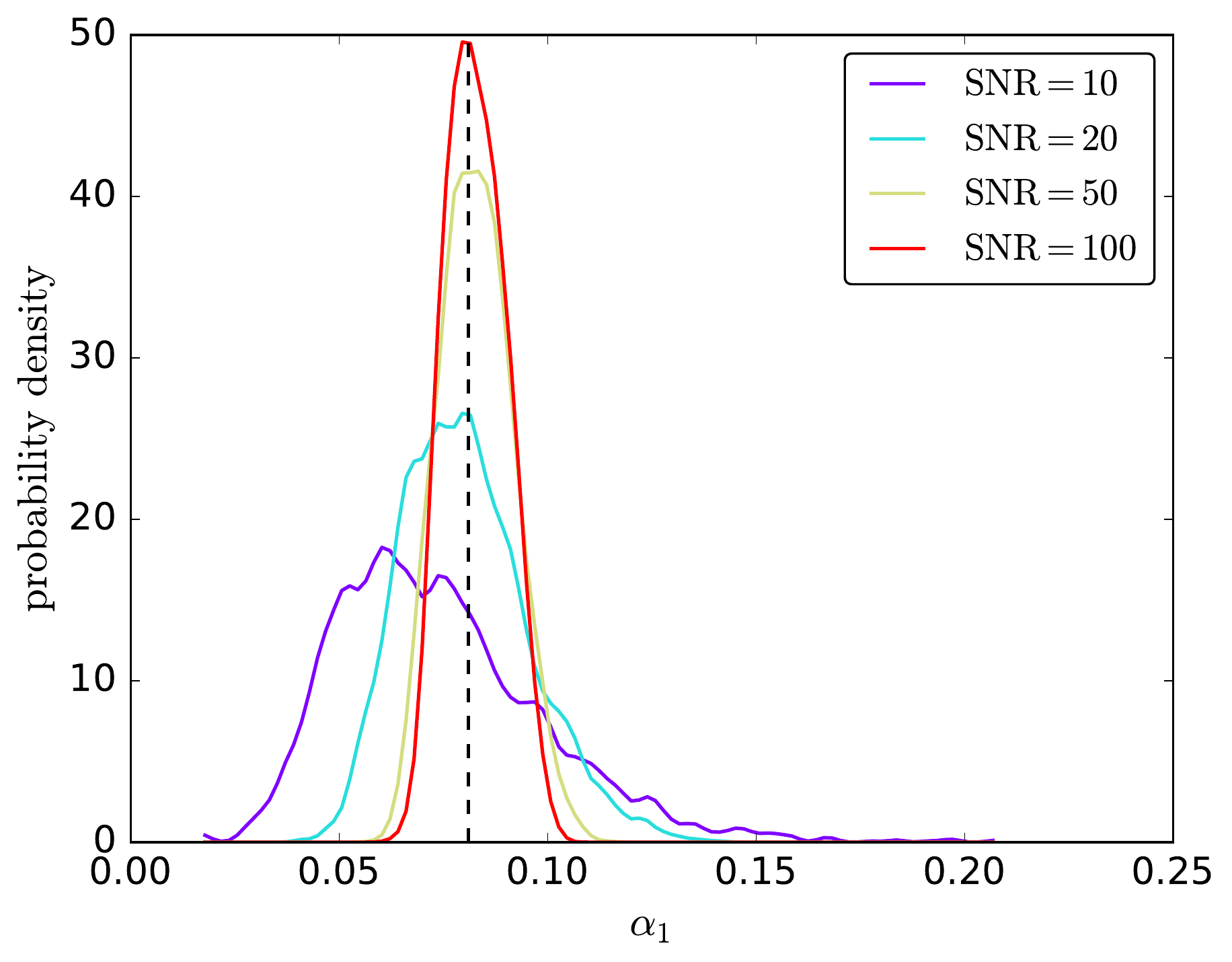}\\
  \includegraphics[width=0.32\textwidth]{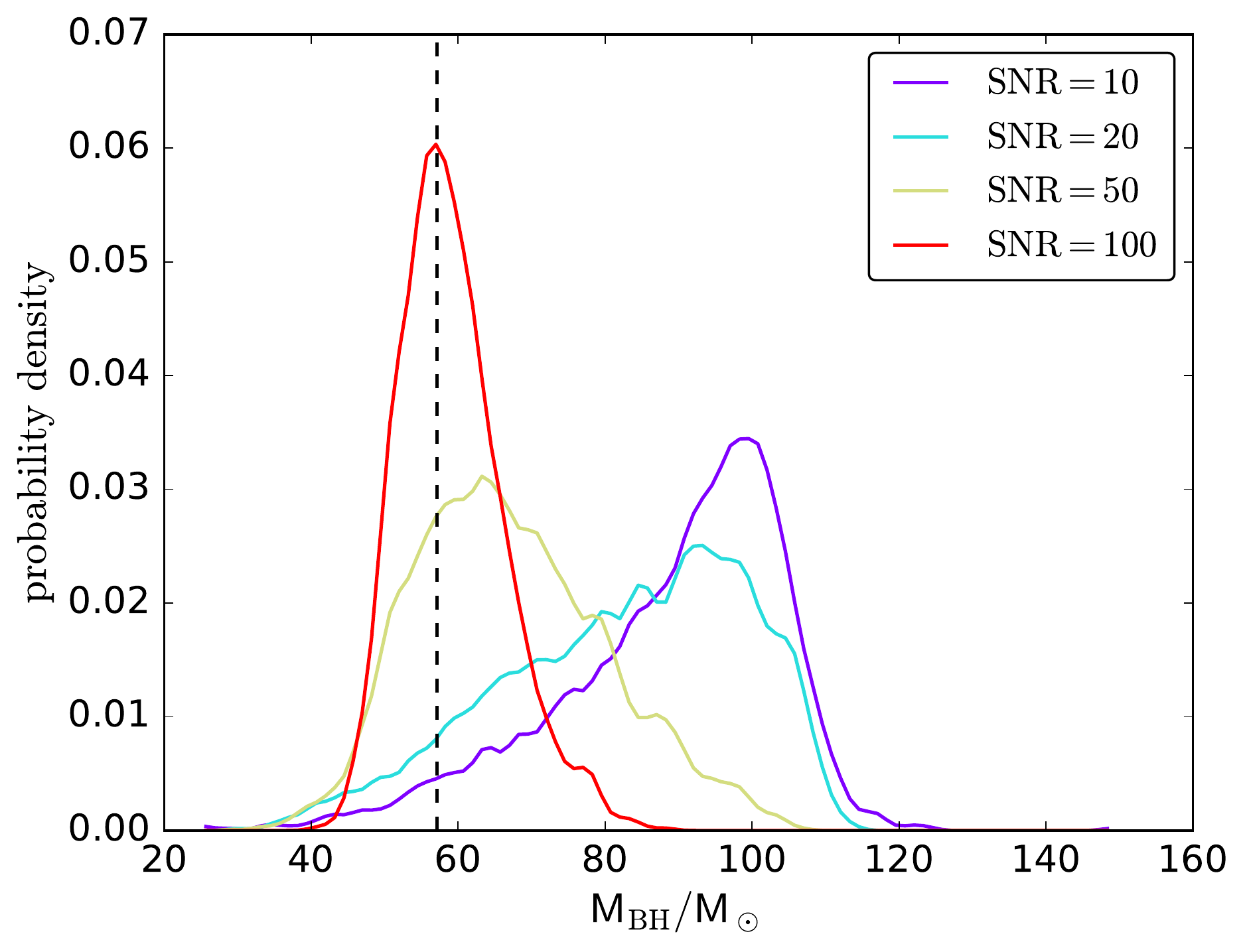}
    \hspace{10mm}
 \includegraphics[width=0.32\textwidth]{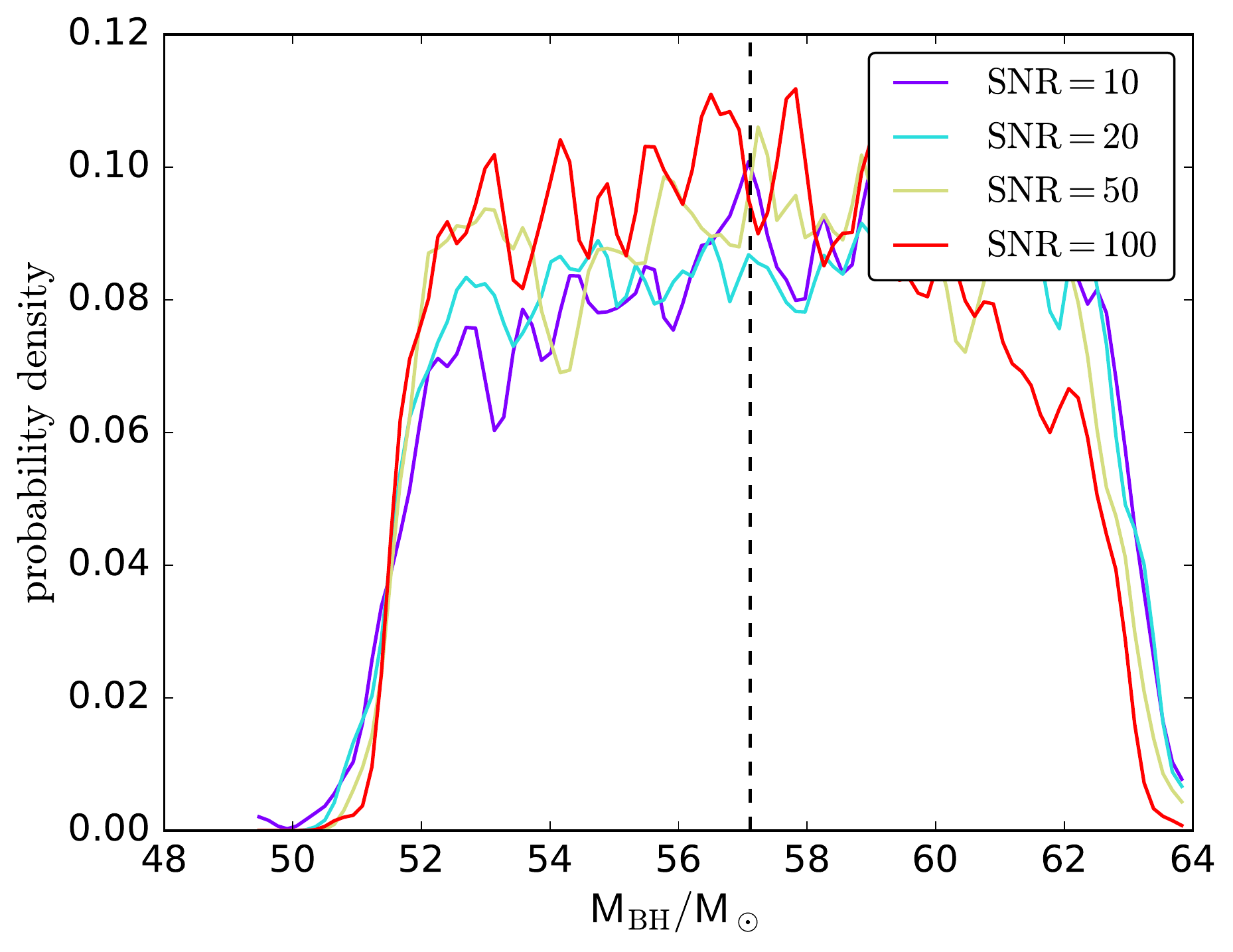}
\end{tabular}
\caption{\label{fig:post_GE2} Measurability of the fundamental QNM frequency $\omega_{1}$ (top), 
inverse damping time $\alpha_{1}$ (middle) and final black hole mass $M_{\rm BH}$ for the 
GW150914-like dataset SXS:BBH:0305. The left panels show posterior distributions probabilities
for the GE case with post-merger SNR=10, 20, 50 and 100. The right panels refer to the CO
case. The vertical line in each panel indicates the correct value of the measured parameter.}
\end{center}
\end{figure*}

As a proof-of-principle, we investigated the accuracy of our template in a simplified,
but realistic, scenario. In all cases we considered, and that are documented below in
Sec.\ref{ss:results}, we used our template for the inference of the physical parameters
of a BBH from the post-merger part of the signal alone. We operate in the context of
Bayesian inference; given the time series output of a detector $d(t)$, we model it as 
\begin{align}
d(t) = h(t;\theta)+n(t)\,,
\end{align}
where $n(t)$ is the detector noise time series and $h(t;\theta)$ is the GW signal
depending on a set of physical parameters $\theta$.
Given the time series $d(t)$ and a waveform model $h(t;\theta)$, our purpose is
to compute the posterior probability distribution for $\theta$. 
To do so, we apply Bayes' theorem:
\begin{align}
p(\theta|d(t),I) = p(\theta|I)\frac{p(d(t)|\theta,I)}{p(d(t)|I)}\,,
\end{align}
where we introduced the prior probability density $p(\theta|I)$, the likelihood
function $\mathcal{L}\equiv p(d(t)|\theta,I)$ and the evidence $p(d(t)|I)$. 
In all terms, we indicate with $I$ whatever background information is relevant to the inference in question.
Since our template discontinuously passes from zero amplitude $A$ for $t < t_0$ to $A\neq 0$ for $t\geq t_0$, we 
find more convenient to perform the analysis in the time domain rather than in the frequency
domain as it is done in most of the literature. Because of the discontinuity, the Fourier transform
of our template would be contaminated by undesirable Gibbs phenomena which would make the inverse transform
not consistent with the sharp time window that defines our template. This could be cured by the convolution in
the frequency domain with the Fourier transform of a square window. 
However, we find simpler to perform the analysis directly in the time domain. For clarity,
and also because this problem is typically reviewed in its frequency domain formulation,
we will briefly go through the fundamentals of the statistical properties of the noise that,
ultimately, are solely responsible for the specific functional form of the likelihood function $\mathcal{L}$. 
We assume, as customary, that the noise is described as a zero-mean wide-sense stationary
Gaussian process. The probability distribution of any given noise realisation at some
countable set of sampling times $t_1,\ldots,t_k$ is thus given by
\begin{align}
p(n_1,\ldots,n_k|I) &\propto e^{-\frac{1}{2} \sum_{i,j} C^{ij}n_i n_j}\,,\\
n_i&\equiv n(t_i)\,,
\end{align}
where ${C^{ij}}$ is the covariance matrix, defined by the stochastic process auto-covariance function $C(t_i,t_j)$:
\begin{align}
C^{ij}\equiv C(t_i,t_j)\equiv <n(t_j)n(t_j)>\,.
\end{align}
Since the noise process is assumed to be wide-sense stationary, we can rewrite 
\begin{align}
C(t_i-t_j,0)\equiv C(\tau) = <n(t)n(t+\tau)>\,.
\end{align}
The auto-covariance function $C(\tau)$ can be learnt from the data in a rather similar fashion
to the Welch method for the estimation of the Power Spectral Density. However, for simplicity,
we are going to assume that the noise process is white, thus $C(\tau) = \sigma^2 \delta(\tau)$.
The covariance matrix ${C^{ij}}$ is therefore diagonal and the probability distribution for the
noise then simplifies to a product of one-dimensional Gaussian distributions
\begin{align}
p(n_1,\ldots,n_k|I) \propto e^{-\frac{1}{2} \sum_{i} \left(\frac{n_i}{\sigma}\right)^2}\,,
\end{align}
which is the functional form for the likelihood $\mathcal{L}$ that we will adopt throughout the rest of the paper. 

\subsection{Simulation set up}
Since the waveform of Eq.~\eqref{eq:wf} depends on several parameters which allow for many possible
degrees of freedom, we consider two set ups:
\begin{enumerate}
\item[GE]: The most generic (GE) case in which no relations among parameters are considered;
\item[CO]:  Partially constrained (CO) case in which we assume that an estimate for the masses 
and spins of the progenitors exists and can be used to impose suitable priors when analyzing 
the post-merger part of signal with the template of Eq.~\eqref{eq:wf}; 
\end{enumerate} 
For all cases, we fix the noise standard deviation $\sigma = 5\times 10^{-22}$ and consider post-merger
signal to noise ratios (SNR) of 10, 20, 50 and 100 by varying the distance $R$ to the source.
Finally, we always consider the source to be optimally oriented and consider only one detector.
We ignore the complications arising from projecting the signal onto the detector tensor since we
are not interested in the inference of any extrinsic parameters like sky position or orientation. 
The set of parameters we consider in each case are:
\begin{enumerate}
\item[GE]: initial masses $(m_1,m_{2})$, the (dimensionless) components of the spins orthogonal
           to the plane of the orbit $(s_{\bot,1},s_{\bot,2})$ (we adopt from now on the notation that
           $s_{\bot,1}$ and $s_{\bot,2}$  indicate the {\it measured} Êvalues of $\chi_{1}$ and $\chi_{2}$), 
           the real, $\alpha_{1}$, and imaginary, $\omega_{1}$, parts of the fundamental mode 
           complex frequency $\sigma_{1}=\alpha_{1}+{\rm i}\omega_{1}$ and the final BH mass $M_{\rm BH}$
\item[CO]: the parameters are the same as above, but with the distinction that the values of 
                  $(m_{1},m_{2})$ and $(s_{\bot,1},s_{\bot,2})$ have a restricted prior.
\end{enumerate} 
In addition to the aforementioned parameters, we always estimate the phase and
time of merger $\phi_0$ and $t_0$ as well as the luminosity distance $R$.

For both GE and CO cases, we analyze all waveforms listed in Table~\ref{tab:NRtest}.
For simplicity, the total initial mass is always fixed to be $60 M_\odot$ when
simulating the detector time series. Finally, we consider a zero-noise realization.
The data are analysed using a dedicated Nested Sampling  algorithm~\cite{cpnest}
similar to Ref.~\cite{LALInference}.

\subsection{Results for GW150914-like dataset}
\label{ss:results}
We start by illustrating our findings through the GW150914-like system SXS:BBH:0305,
that corresponds to $(q,\chi_1,\chi_2)\simeq (1.22,+0.33,-0.44)$. We show posterior
distributions for the general case GE in Figs.~\ref{fig:post_GE_wf} (waveform reconstruction)
and in Figs.~\ref{fig:post_GE1}-\ref{fig:post_GE2} for the posterior probability distributions 
of the various physical parameters. For all waveforms in Table~\ref{tab:NRtest}, we compute 
posterior probability distributions for all the parameters listed in the previous Section 
and corresponding to the cases GE and CO.
Figure~\ref{fig:post_GE_wf} shows the $90\%$ confidence waveform recovered by our analysis 
as a function of the post-merger SNR going from 10 (top panel) to 100 (bottom panel). 
The left column is pertinent to the GE case while the right column to the CO case. 
In all cases one notices the shrinking of the $90\%$ credible region with increasing SNR.
However, we also note two interesting facts: (i) the general behavior of the
waveform is always well recovered, even in the lowest SNR case; (ii) there is not much
difference between the GE and the CO case, indicating that, at least for systems
like SXS:BBH:0305, the main factors determining the actual shape of the waveform are
not much the initial properties of the coalescing system,
but rather the properties of the final BH. This is further exemplified by looking at the details of
the probability density functions (PDFs) given in Figs.~\ref{fig:post_GE1} and~\ref{fig:post_GE2}. 
In the GE case, we see that the component masses $m_1$ and $m_2$ are largely unconstrained for
SNR~$<50$ and begin to be well measured ($O(30\%)$ and $O(20\%)$) for SNR = 50 and 100, respectively.
At the same time, the spin magnitudes $s_{\bot,1}$ and $s_{\bot,2}$ are never constrained. 
We note a similar behavior for the ringdown part of the waveform.
The QNM fundamental frequency $\omega_1$ can be constrained only when the
post-merger SNR is $\geq 50$ while the (inverse) damping time $\alpha_1$ is always measured,
improving to $O(25\%)$ and $O(15\%)$ for SNR = 50 and 100, respectively. Finally, we find that  
also the final black hole mass $M_{\mathrm{BH}}$ is measurable only for post-merger SNR $\geq 50$.
In the CO case, the situation is remarkably similar. Apart from the obvious fact that the
component masses are determined by their prior, no measurement of the component spins is
possible, at least for the SNRs considered in this work. A main difference with the GE case
is the accuracy with which the complex ringdown frequency can be determined. Remarkably,
the posterior for $\omega_1$ does not seem to be affected the SNR considered which suggests
that its posterior is mainly determined by correlations in the $m_1,m_2,\omega_1$ sub-parameter space;
$\alpha_1$ instead is determined with $30\%, 17\%, 10\%$, and $8\%$ accuracy for
SNR = $10,20,50,100$. Finally and not surprisingly also $M_{\mathrm{BH}}$ is always
well determined, independently of the post-merger SNR.

\begin{figure}[t]
\begin{center}
\includegraphics[width=0.40\textwidth]{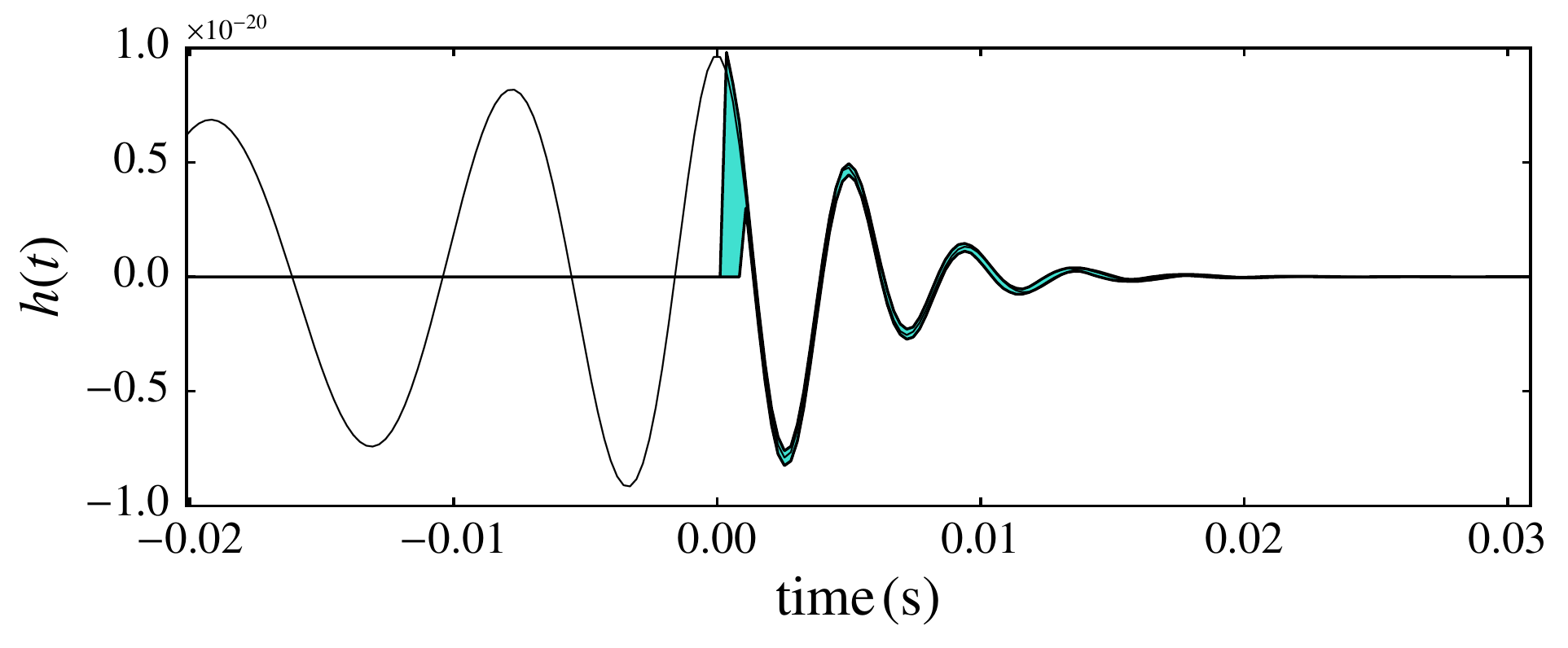}\\
\includegraphics[width=0.41\textwidth]{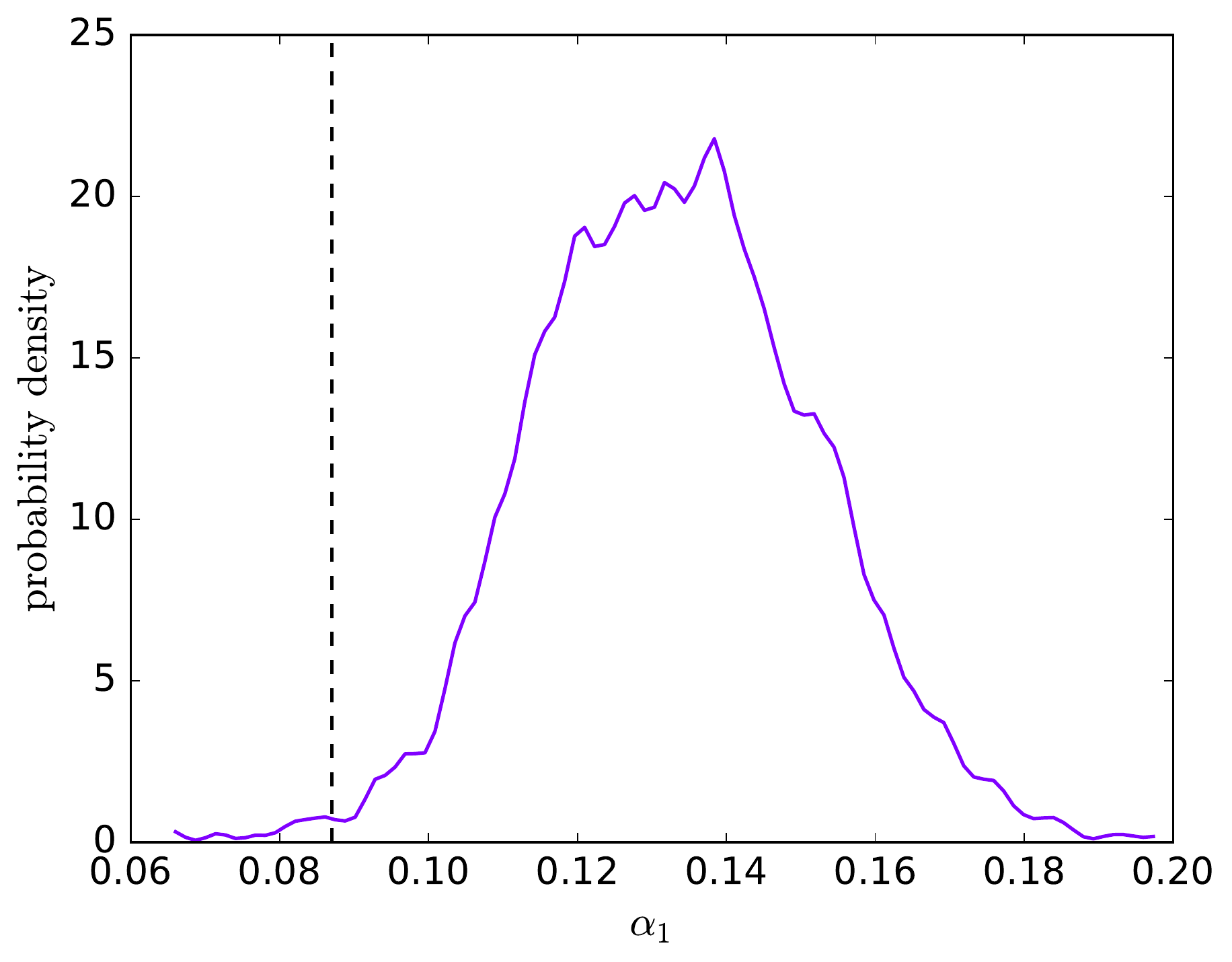}\\
\includegraphics[width=0.41\textwidth]{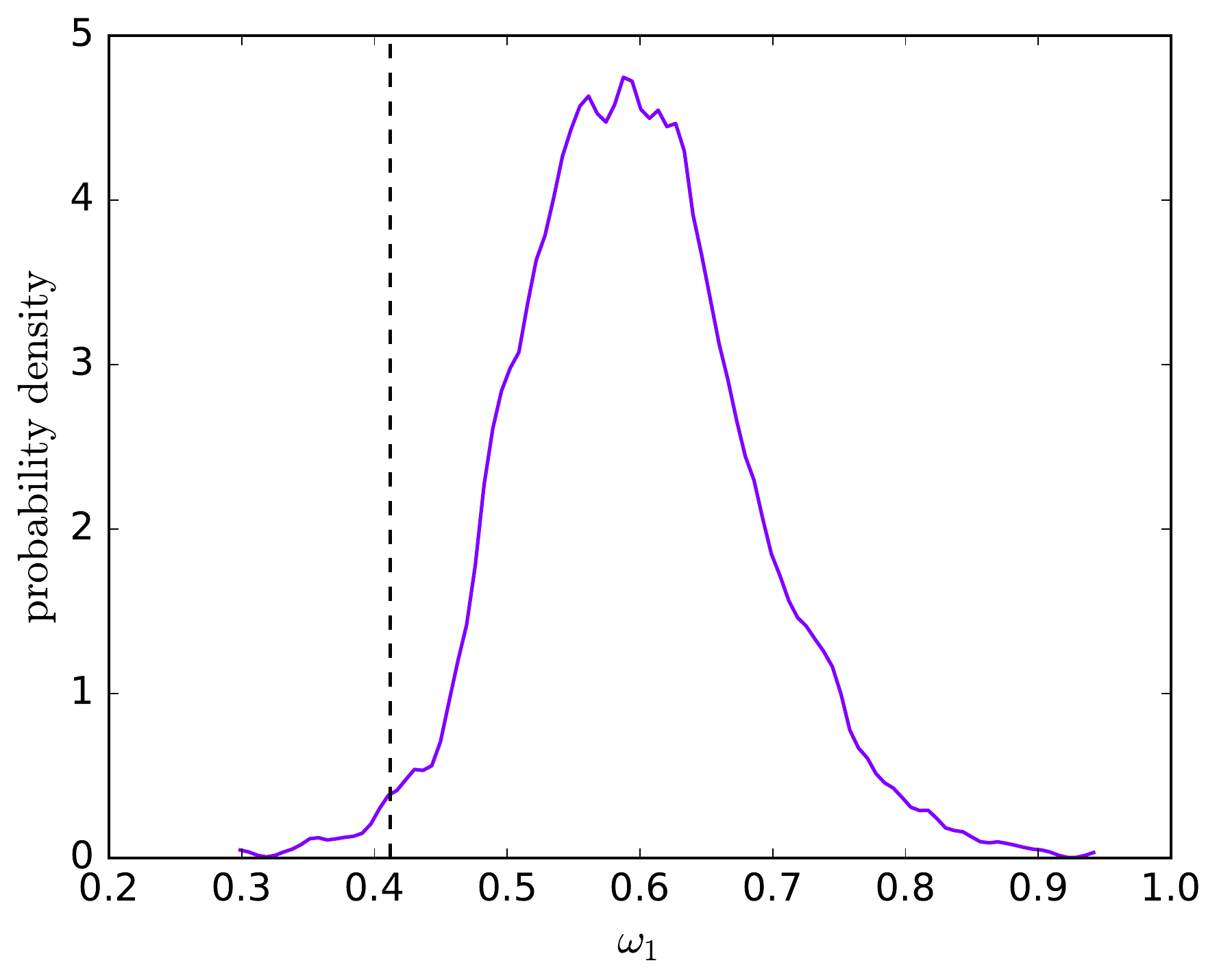}
\caption{\label{fig:wf_qnm_q10}Top panel: $90\%$ credible waveform for SXS:BBH:0185. Middle and bottom
  panels: posterior distributions probabilities for the general unconstrainded case for the fundamental
  QNM complex frequency for ${\rm SNR} = 50$. The waveform reconstructed is well consistent with the underlying
  injected NR waveform, though the QNMs frequency show significan bias (the vertical line indicates the
  exact value).}
\end{center}
\end{figure}

\subsection{Results for the complete dataset}
\label{sec:discussion}
We then use the template to extract the parameters from all waveform listed in Table~\ref{tab:NRtest} 
and we report  only on the (most demanding) GE case, i.e. no priors for the components masses and 
spins are assumed.  The measured (complex) frequency of the fundamental QNM of the final BH
as well as its mass $M_{\rm BH}$ are listed in the last columns of Table~\ref{tab:NRtest},
either with post merger ${\rm SNR=10}$ and ${\rm SNR=50}$. Comparing the measured frequency with the exact value
(third column in the table) one sees, on average, a good consistency between the measured
and expect values. There are however a few exception, where the measured frequency looks
biassed. Interestingly, we found that this effects persists also for ${\rm SNR>20}$,
indicating that the template does show some systematics.
For example, this is the case for $(9.99,0,0)$ and $(5,-0.90,0)$.
The former case is illustrated in Fig.~\ref{fig:wf_qnm_q10}. The top panel shows the reconstructed
$90\%$ waveform compared to the injected one for ${\rm SNR} = 50$. This waveform is consistent
with the expectations, suggesting that the functional ansatz we adopted are indeed appropriate
to represent the true BBH waveform. However, the middle and lower panel in Fig.~\ref{fig:wf_qnm_q10}
show how biased is the complex frequency of the fundamental QNM.
We speculate it to be due to the fact that this particular dataset stands
{\it outside} the domain of calibration of the template (we recall that the largest mass-ratio
we include is $q=8$) and that the extrapolation there is then not very accurate. To test whether
this hypothesis is correct one will need to incorporate in the template (i) NR datasets with larger
mass ratios (e.g., see e.g. Ref.~\cite{Khan:2015jqa}, that computed NR data with $q=18$) and (ii)
large-mass ratio data~\cite{Harms:2014dqa}, though, as mentioned above, this will require, to be
accurate, a new ansatz for the fitting template.
For the case $(5,-0.90,0)$, the bias is probably due to the fact that the template is unable to
account for the mixing between prograde and retrograde modes, an effect that, as mentioned already
above, is evident when inspecting the phase difference in Fig.~\ref{fig:figFit}. Improving the
post-merger waveform model with the primary goal of not having this biases will be our primary
aim for future work.

\subsection{Testing the second law of BH dynamics}
The specific nature of our template allows us to determine the initial masses
and spins as well as the final mass and spin (the latter via the fundamental QNM)
in an essentially independent fashion. In 1971 and 1972, Hawking proposed and
proved the so-called ``area theorem''
or ``second law of black hole dynamics''~\cite{Hawking:1971vc,Bardeen:1973gs}: 
\begin{quotation}
when black holes collide the sum of the surface areas of all black
holes involved can never decrease~\cite{Misner:1974qy}. 
\end{quotation}
By comparing the component masses and spins with the final mass and spin
inferred from our analysis, we can compare the sum of the initial Kerr
areas of the black hole binary with the Kerr area of the final black
hole, without assuming a full coherent inspiral-merger-ringdown
waveform that, being built on General Relativity, naturally satisfies
the area theorem. We compare the posterior distributions on the Kerr
areas as follows. Define $A_1+A_2 = X$ the sum of the initial areas
and $A_f$ is the area of the final black hole. 
For any Kerr black hole:
\begin{align}
r_s &= \frac{G M}{c^2},\\
A &= 4\pi(r_+^2+(a r_s)^2),\\
r_+ &= r_s\left[1+\sqrt{1-a^2},\right]
\end{align} 
with $M$ the black hole mass and $a$ the dimensionless spin parameter.
Our purpose is to calculate $p(A_f\geq A_1+A_2)$. If $f(A_f)$ is the
probability density function for $A_f$ and $g(A_1+A_2)$ is the one for
the sum of the initial areas, we define a new
variable $z = A_f-(A_1+A_2)$ and calculate 
\begin{align}
p(z) = \int dA_f f(A_f)g(A_f-z),
\end{align}
from which it is possible to compute the probability that the final
area is greater than the sum of the initial. For the GW150914-like dataset
SXS:BBH:0305, we find probabilities $(0.74, 0.74, 0.78,0.85)$,
respectively for ${\rm SNR}=(10,20,50,100)$, that the final area is larger
than the initial one. This preliminary result illustrates the feasibility
of this measurement. We will report an an extensive investigation of
this aspect in a future publication.

\section{Concluding remarks}
\label{sec:conclusions}
We proposed a post-merger, time-domain, waveform template of the form
\be
\label{eq:hTemp}
h(\tau)=e^{-\sigma_1\tau-{\rm i}\phi_0}\bar{h}(\tau;\nu,\ha_0),
\ee 
where $\tau=(t-t_0)/M_{\rm BH}$, $\ha_0=\ta_1+\ta_2$ 
and $\nu=m_1 m_2/(m_1+m_2)^2$. 
The complex fundamental QNM frequency 
$\sigma_1\equiv \alpha_1+{\rm i}\omega_1$ is itself a free parameter. 
We have also shown that this waveform is accurate and does provide 
a useful template to analyze the post-merger signal of systems like GW150914 
that depends on the physical parameters of the system. In particular, 
modelling the post-merger part of the coalescence, the analysis presented 
in Ref.~\cite{TheLIGOScientific:2016src}, could be improved by either  
fixing the initial time $t_0$ much closer to the merger time $t_{\rm M}$, 
thus recovering more SNR, or marginalizing over the initial time $t_0$ 
so to avoid arbitrary choices for the beginning of the ringdown. 
Our template also provides a way of improving the inspiral-merger-ringdown 
consistency test~\cite{TheLIGOScientific:2016src}. The current test relies on 
the comparison of the posteriors reconstructed assuming the \emph{same} 
waveform model in different frequency regimes. Our template provides an 
independent way of extracting physical information about the BBH system 
from the post-merger phase only. In principle, our template also gives a 
means to extract the \emph{full} information about the original binary from 
a detailed analysis of the post-merger/ringdown signal. The feasibility and 
SNR requirements of this are currently being explored. Finally, the functional 
representation of the post-merger part given by Eq.~\eqref{eq:hTemp} 
is easily generalized to allow for more freedom in the waveform. 
Some of the physical parameters entering in the vector $Y$ could be 
treated as free parameters and thus inferred from the data
rather than being extracted from the NR simulations. 
For instance, in case of $\alpha_1$ and $\alpha_2$,
i.e. the inverse damping time of the fundamental QNM and of the first overtone, one could
relax the constraint $\alpha_{21}=\alpha_2-\alpha_1$ and keeping $\alpha_1$
as a free parameter in $\bar{h}$. Rather than Eq.~\eqref{eq:hTemp}, 
one would use a post-merger template of the form
\be
\label{eq:halpha}
h(\tau)=e^{-(\alpha_1+{\rm i}\omega_1)\tau-{\rm i}\phi_0}\bar{h}(\tau;\nu,\ha_0,\alpha_1,\alpha_2),
\ee
where $(\alpha_1,\alpha_2,\omega_1)$ are all considered free parameters to be
inferred from the experimental data. Measuring $\alpha_2$, one could 
setup a (partial) test of the general-relativistic no-hair 
theorem~\cite{Dreyer:2003bv,Berti:2005ys,Berti:2007zu,Gossan:2011ha,Meidam:2014jpa} 
by estimating the consistency between $(\omega_1,\alpha_1)$ and $\alpha_2$.

\begin{acknowledgments}
AN thanks Chris~Van~Den~Broeck for answering a question as well as  
for the clarifying follow up discussion that eventually prompted this work;
and Badri~Krishnan, Andrew~Lundgren and Miriam~Cabero for
useful discussion and comments on the very first draft of this manuscript. 
AN is also grateful  to Thibault Damour for strongly suggesting to develop 
this idea in a paper,  and to Philipp Fleig for a careful reading of the 
very first version of the manuscript. Finally, we would like to thank
Mark~Hannam and Sascha~Husa for lending us their, nonpublic, BAM waveform
to do the few preliminary tests of the template discussed in the text.

We also gratefully acknowledge correspondence with Alejandro Boh\'e aimed
at clarifying the origin of the discrepancy between the reproduction of
our results shown in Appendix~B of the first version of Ref.~\cite{Bohe:2016gbl}
and what presented here. Once the issue was clarified, full consistency
with our results was found. See in particular footnote~3 in the main text.

This work is dedicated to the memory of 
Ulysses, that for more than ten years, silently, shared several 
important moments and ``{\it visse per seguir virtute e canescienza}''.
\end{acknowledgments}

\bibliography{refs20170112}      

\end{document}